\documentclass[showpacs,twocolumn,amsmath,amssymb,prx,superscriptaddress,longbibliography]{revtex4-1}

\usepackage[unicode]{hyperref}
\hypersetup{
   unicode=true,          
   plainpages=false,
   colorlinks=true,       
   citecolor=blue,        
}
\usepackage{color}
\usepackage{graphicx}
\usepackage{amsmath}
\usepackage{amssymb}
\usepackage{bm}
\usepackage{url}
\usepackage{tikz}
\usepackage{array}
\newcolumntype {s}[1]{@{\hspace{#1}}} 
\usepackage{colordvi}

\graphicspath{{.}{./EPS/}}

\makeatletter
\g@addto@macro\normalsize{%
  \setlength\abovedisplayskip{5pt}
  \setlength\belowdisplayskip{5pt}
  \setlength\abovedisplayshortskip{5pt}
  \setlength\belowdisplayshortskip{5pt}
}
\makeatother

\newcommand{\rme}{\mathrm{e}}
\newcommand{\rmi}{\mathrm{i}}

\newcommand{\field}[1]{\mathbb{#1}} 
\renewcommand{\d}{\ensuremath{\mathrm{d}}}

\begin{document}

\title{Exactly solvable model for a solitonic vortex in a compressible superfluid}

\author{L. A. Toikka}
\email{L.A.Toikka@massey.ac.nz}
\author{J. Brand}
\email{j.brand@massey.ac.nz}
\affiliation{Dodd-Walls Centre for Photonic and Quantum Technologies, Centre for Theoretical Chemistry and Physics, and New Zealand Institute for
Advanced Study, Massey University, Private Bag 102904 NSMC, Auckland 0745, New
Zealand}

\begin{abstract}
Vortex motion is a complex problem due to the interplay between the short-range physics at the vortex core level and the long-range hydrodynamical effects. Here we show that the hydrodynamic equations of vortex motion in a compressible superfluid can be solved exactly in a model ``slab'' geometry. Starting from an exact solution for an incompressible fluid, the hydrodynamic equations are solved with a series expansion in a small tuneable parameter provided by the ratio of the healing length, characterizing the vortex cores, to the slab width. The key dynamical properties of the vortex, the inertial and physical masses, are well defined and renormalizable. They are calculated at leading order beyond the logarithmic accuracy that has limited previous approaches. Our results provide a solid framework for further detailed study of the vortex mass and vortex forces in strongly-correlated and exotic superfluids. The proposed geometry can be realised in quantum-gas experiments where high-precision measurements of vortex mass parameters are feasible.
\end{abstract}


\date{\today}
\maketitle

\section{Introduction} \label{sec:intro}
The quantisation of vortices in superfluids is one of the most prominent quantum phenomena of condensed matter physics. Yet the dynamical properties of vortices are  poorly understood, with the mass \cite{Suhl1965,Baym1983,Duan1992}
and applicable forces \cite{Iordansky1964a,Thouless1996,Kopnin2002}
still the subject of controversies after decades of research \cite{Thouless2007,Thompson2012,Golubchik2012,Bevan1997,Volovik2013,Sonin2013}.
Recent progress in ultra-cold atomic gases has demonstrated that the dynamics of individual vortices can be observed in defect-free superfluids near zero temperature \cite{Anderson2000,Freilich2010,Yefsah,Ku2014,Serafini2015}. In particular the determination of the inertial to physical mass ratio  from the observed oscillation frequency of a solitonic vortex in a trapped unitary Fermi gas in Ref.~\cite{Yefsah} has demonstrated that, in principle, high precision measurements of the dynamical properties are feasible \footnote{Initially the observed defects were interpreted as dark solitons \cite{Yefsah} but later correctly identified as vortices \cite{Ku2014}.}.
The ability to derive a deeper  understanding of vortex motion from such experiments is currently impeded by the limited precision of available theoretical predictions. For harmonically trapped quantum gases
 \cite{fetter01:vortices,Koens2012,Pitaevskii2013,Ku2014} these are typically performed at logarithmic accuracy, i.e.~up to an undetermined factor that varies only slowly (logarithmically) with system parameters.
In addition to the long-range hydrodynamic laws governing the vortex motion, one can also expect the short-range, mesoscopic structure of the vortex core and its interaction with the fluid to influence the precise value of the vortex mass and other dynamical properties. This is particularly interesting for strongly-correlated quantum fluids such as superfluid helium \cite{leggett2006quantum} where no quantitative microscopic theory exists. In Fermi superfluids such as neutron matter \cite{Archibald2013} and the unitary Fermi gas \cite{Zwerger2012} one may further expect the Kopnin  \cite{Kopnin2002} and Iordanskii forces \cite{Iordansky1964a} to become relevant even though their effects may easily be dwarfed by dominant contributions from long-range hydrodynamics including the Magnus force \cite{Saffman1995}. Hence it is important to have precise control and understanding of the long-range hydrodynamic effects on the vortex motion.
In this work we propose a simple slab geometry that provides this precise control and solve the hydrodynamic Euler equations for the vortex motion by a series expansion, where each term can be computed exactly.

Thouless and Anglin concluded from a theoretical study that the vortex mass is poorly defined as its value depends on the process of measurement  \cite{Thouless2007}. While this remains true in general, we demonstrate  that the vortex mass can in fact be properly defined if an appropriate context is provided. Here this is achieved by specifying the geometry of a long but transversely confined channel, or slab. A vortex line perpendicular to the channel's long axis is known as a \emph{solitonic vortex} \cite{brand01a,Brand2002,Komineas2003}. The concept was introduced in \cite{brand01a} in order to describe localised nonlinear wave excitations with vortex structure that travel at sub-sonic velocities along the channel \cite{Komineas2003}. Solitonic vortices have recently been observed as long-lived products from the decay of dark solitons \cite{PhysRevLett.116.045304,Ku2014,Becker2013} and as spontaneously formed defects during the Bose-Einstein condensation transition \cite{Lamporesi2013,Donadello2014}. 
The solitonic vortex appears as a decay product of other families of non-linear localised excitations and is, in fact, the only known stable and slow-moving solitary wave excitation in channels that are sufficiently wide to host a vortex \cite{1367-2630-17-12-125013}.

\begin{figure*}[t]
\vspace{-0.5cm}
\begin{tikzpicture}[baseline]
\xdefinecolor{darkgreen}{RGB}{175, 50, 36}
\pgfmathsetmacro{\xbase}{-4.0}
\pgfmathsetmacro{\ybase}{0}
\pgfmathsetmacro{\xoffset}{0}
\pgfmathsetmacro{\xoffsetl}{9.0}
\pgfmathsetmacro{\xoffsetleg}{3.9}
\pgfmathsetmacro{\yoffsetcont}{0.35}
\pgfmathsetmacro{\yoffset}{-2.5}
\pgfmathsetmacro{\yoffsetl}{0.15}
\pgfmathsetmacro{\lscale}{0.0115}
\pgfmathsetmacro{\yb}{1.5}

\node[scale=0.95] at (0,0) {\includegraphics[width=0.4\textwidth]{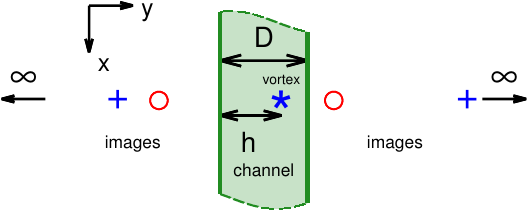}};
\node at (\xbase,1.5) {(a)};

\node[scale=0.95] at (0.3,\yb*\yoffset) {\includegraphics[width=0.49\textwidth]{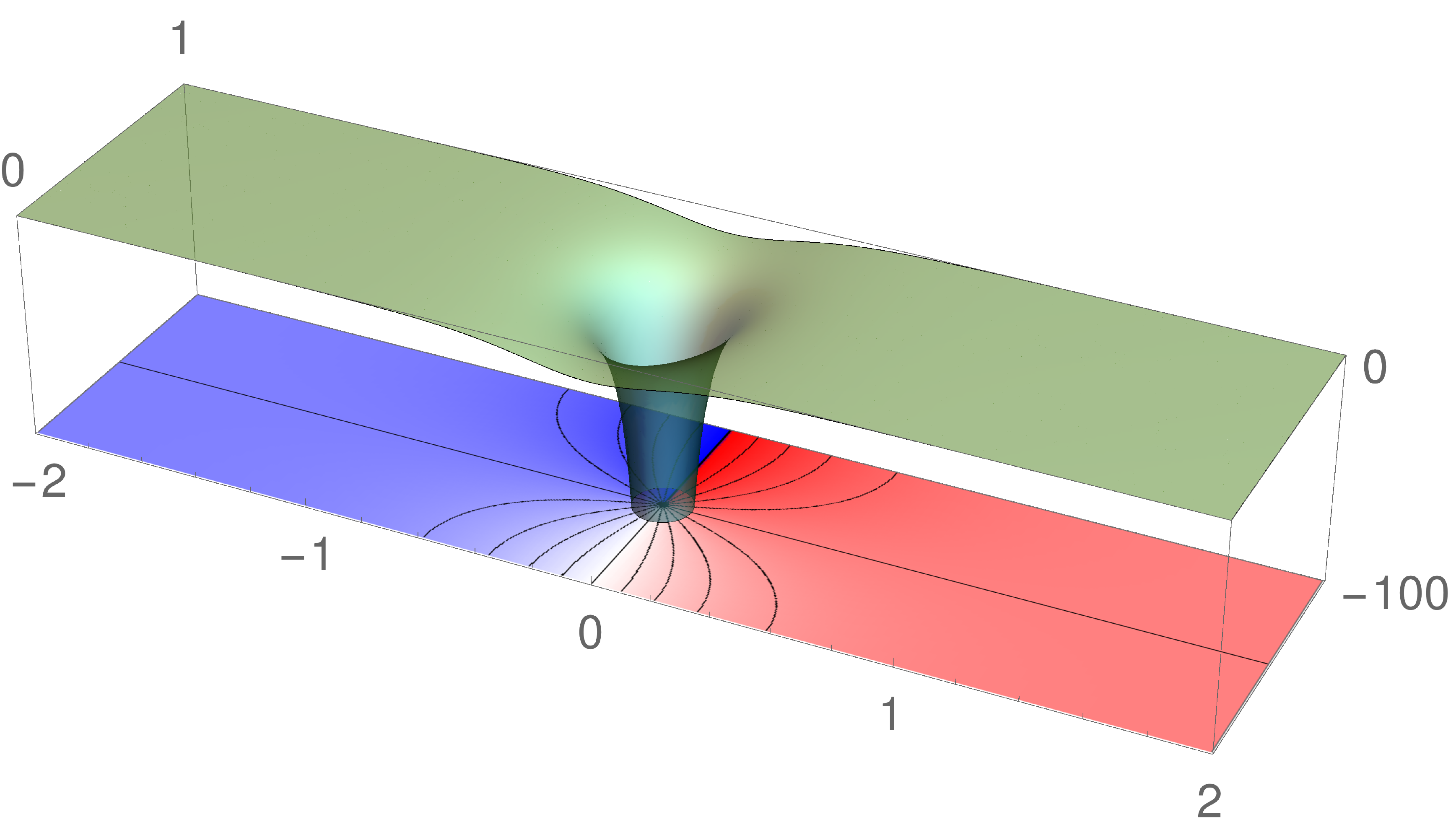}};
\node at (\xbase+8.5,-0.3+\yb*\yoffset) {$\frac{n_1}{n_0 q^2 \xi^2/D^2}$};
\node at (\xbase+3.5,-1.7+\yb*\yoffset) {$x/D$};
\node at (\xbase+0.5,2.0+\yb*\yoffset) {$y/D$};
\node at (\xbase+6.5,1.3+\yb*\yoffset) {\color{darkgreen}{1st order density $n_1$}};
\node at (\xbase+3.0,-2+\yb*\yoffset) {\color{blue}{Incompressible phase $S_0/q$}};
\node at (\xbase,\yoffsetcont+\yb
*\yoffset+1.5+\yoffsetl) {(b)};
\node[scale=0.87,rotate=-90] at (\xbase+1.3,-1.5+\yb*\yoffset) {\includegraphics[width=\lscale\textwidth]{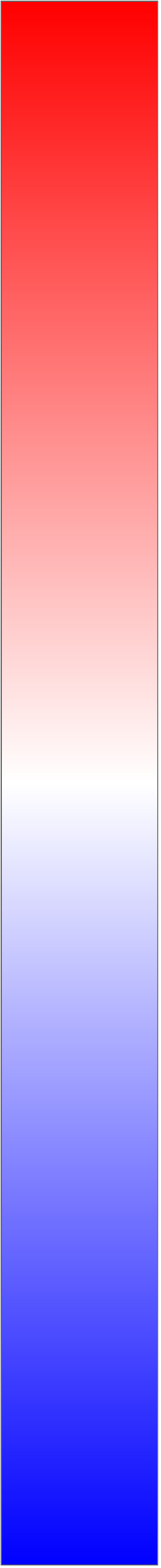}};
\node[scale=0.8] at (\xbase+0.4,-1.3+\yb*\yoffset) {$-\pi$};
\node[scale=0.8] at (\xbase+1.3,-1.27+\yb*\yoffset) {$0$};
\node[scale=0.8] at (\xbase+2.2,-1.3+\yb*\yoffset) {$\pi$};

\node[scale=0.95] at (\xoffsetl,\yoffsetcont) {\includegraphics[width=0.47\textwidth]{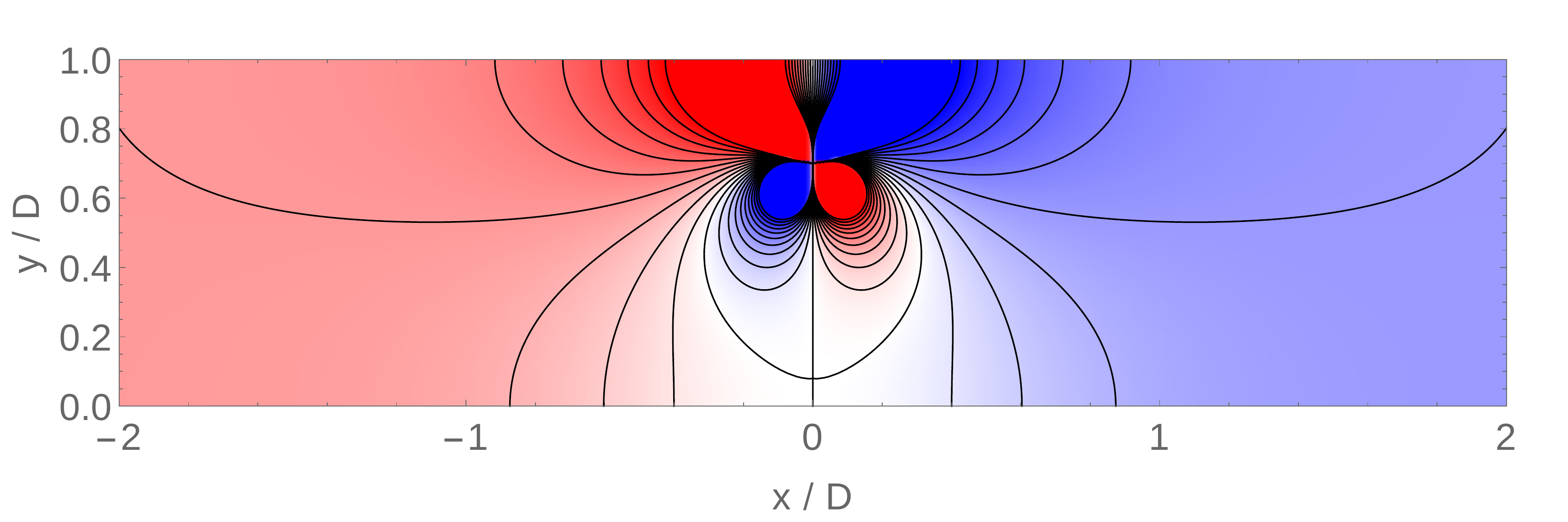}};
\node[scale=0.87] at (\xoffsetl+\xoffsetleg,\yoffsetcont+\yoffsetl) {\includegraphics[width=\lscale\textwidth]{1l}};
\node[scale=0.8] at (\xoffsetl+4.3,\yoffsetcont-0.9+\yoffsetl) {$-4\pi$};
\node[scale=0.8] at (\xoffsetl+4.2,\yoffsetcont+\yoffsetl) {$0$};
\node[scale=0.8] at (\xoffsetl+4.3,\yoffsetcont+0.9+\yoffsetl) {$4\pi$};
\node at (\xbase+\xoffsetl,\yoffsetcont+1.0+\yoffsetl) {(c)};
\node at (\xbase+\xoffsetl+4.2,\yoffsetcont+1.2+\yoffsetl) {\color{blue}{1st order phase $S_1/(q^3 \xi^2/D^2)$}};

\node[scale=0.95] at (\xoffsetl+0.5,\yoffsetcont+1.7*\yoffset) {\includegraphics[width=0.47\textwidth]{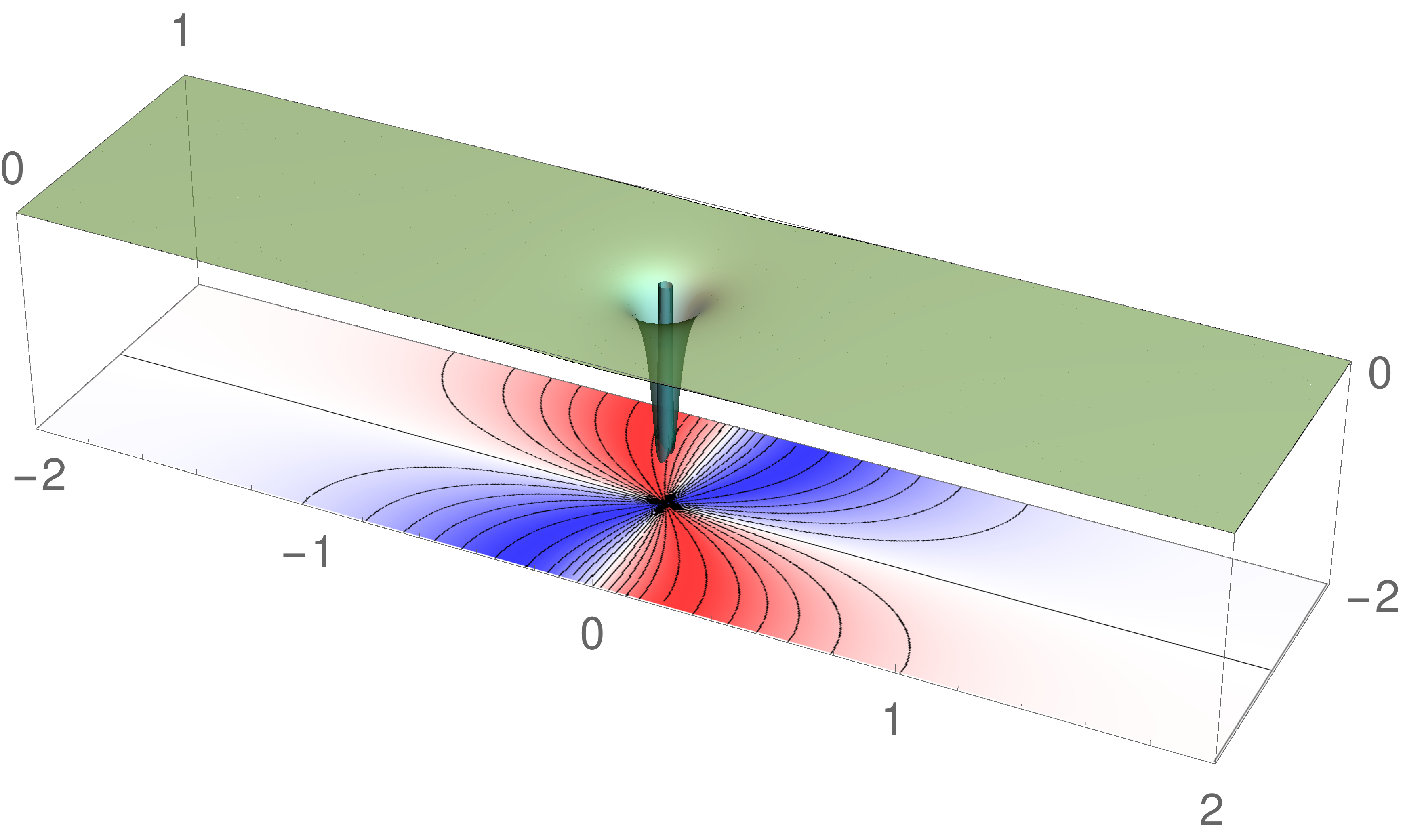}};
\node at (\xbase+\xoffsetl,\yoffsetcont+\yb
*\yoffset+1.5+\yoffsetl) {(d)};
\node at (\xbase+\xoffsetl+8.7,-0.3+\yb*\yoffset) {$\frac{n_1 + n_2}{n_0}$};
\node at (\xbase+\xoffsetl+3.5,-1.7+\yb*\yoffset) {$x/D$};
\node at (\xbase+\xoffsetl+0.8,1.8+\yb*\yoffset) {$y/D$};
\node at (\xbase+\xoffsetl+6.5,1.3+\yb*\yoffset) {\color{darkgreen}{2nd order density $n_1 + n_2  (\mathrm{UFG})$}};
\node at (\xbase+\xoffsetl+5.5,-2.2+\yb*\yoffset) {\color{blue}{1st order phase $\frac{S_1}{q^3\xi^2/D^2}$}};
\node[scale=0.87,rotate=-90] at (\xoffsetl+\xbase+1.3,-1.5+\yb*\yoffset) {\includegraphics[width=\lscale\textwidth]{1l}};
\node[scale=0.8] at (\xoffsetl+\xbase+0.4,-1.3+\yb*\yoffset) {$-\pi$};
\node[scale=0.8] at (\xoffsetl+\xbase+1.3,-1.27+\yb*\yoffset) {$0$};
\node[scale=0.8] at (\xoffsetl+\xbase+2.2,-1.3+\yb*\yoffset) {$\pi$};

\end{tikzpicture}
\caption{\label{fig:1}
\textbf{Solitonic vortex in a compressible superfluid.} (\textbf{a}) \textit{Method of images}: Placement of two infinite families of images of alternating sign, of which the two first images are shown. Pluses (circles) correspond to images with the same (opposite) charge as the real vortex (*). 
(\textbf{b}) The first-order compressible density profile $n_1 = -q^2 n_0 \xi^2 |\nabla S_0 |^2$ relative to the leading-order incompressible constant background $n_0$ together with $S_0$, the incompressible phase profile. Here $h = D/2$. 
(\textbf{c}) Numerical evaluation of the renormalised first-order compressible phase correction $S_1$ with $h/D = 0.7$. 
(\textbf{d}) Same as \textbf{b}, but with the second-order density correction $n_2$, which depends on $S_0$ and $S_1$. For $n_1+n_2$, we have chosen $q = 1$, $\xi/D = 0.05$, and $\gamma = 3/2$, representing the unitary Fermi gas (UFG). Note that $n_1$ is independent of $\gamma$, the polytropic index.} 
\end{figure*}

By virtue of the transverse confinement, the vortex becomes a localised object. Indeed, the exact solutions obtained in this work and visualised in Fig.~\ref{fig:1} show that the associated velocity field and the corresponding  excitation energy density are exponentially localised in the longitudinal direction on the length scale of the transverse dimension. Thus the vortex loses its long-range nature beyond the transverse length scale of the channel. The localisation makes it possible to consider the solitonic vortex as a quasiparticle with a well-defined inertial mass $M^*$ through the framework of Landau quasiparticle dynamics of solitary waves \cite{Konotop2004}. It allows us to derive a Newtonian equation of motion under the condition that the localised quasiparticle moves sufficiently slowly between regions of different background density, to suppress radiation and to conserve the quasiparticle's energy. The effects of a trapping potential are hereby encapsulated in the chemical potential of local equilibrium $\mu(X) = \mu_0- V_\mathrm{trap}(X)$, which is assumed to vary by negligible amounts over the length scale of the solitonic vortex. If $E_\mathrm{s}(\mu,{V})$ denotes the excitation energy associated with a solitonic vortex at chemical potential $\mu$ moving with velocity ${V} = \dot{X}$, then requiring conservation of energy, $\d E_\mathrm{s}(\mu(X),\dot{X})/\d t =0$, leads to
\begin{equation}
\label{eq:Newton}
F = M^* \ddot{X},
\end{equation}
where 
\begin{equation}
M^* = \frac{1}{{V}} \left.\frac{\partial E_\mathrm{s}}{\partial {V}}\right|_\mu
\end{equation}
is the inertial or effective mass. Since the vortex depletes particles from its core, we also have $F= M_\mathrm{ph} g$, a buoyancy-type force with the physical mass
\begin{equation}
\label{eqn:physMassdef}
M_\mathrm{ph} = -m \left.\frac{\partial E_\mathrm{s}}{\partial \mu}\right|_{{V}},
\end{equation}
where $g = m^{-1} \d\mu/\d X$ is a buoyancy parameter with dimension of acceleration, and $m$ the mass of the elementary bosons. While $g$ characterises the environment of the solitonic vortex, the mass parameters $M^*$ and $M_\mathrm{ph}$ are measurable and well-defined intrinsic properties that depend both on the microscopic details of the many-body physics, e.g.\ the vortex core structure, as well as the transverse confinement. In this way the Landau quasiparticle picture of the solitonic vortex provides a framework for defining vortex properties, where additional effects like non-conservative and fluctuating forces can be added later.

Here, we will calculate the mass parameters $M^*$ and $M_\mathrm{ph}$ that arise as a consequence of the superfluid hydrodynamics given by the Euler and continuity equations. These equations  govern the fluid dynamics at length scales large compared to the healing length $\xi$ and other microscopic length scales. The ratio $\xi/D$ of the healing length to the channel width $D$ thus naturally emerges as a small parameter of the theory. The key to finding exact solutions is to choose a geometry where simple analytic solutions are available for the simplified model of an incompressible fluid. From this starting point, compressible corrections are later calculated as a perturbation series.

The paper is organised as follows: The main hydrodynamic theory is laid out in Sec.~\ref{sec:hyd}, starting with the definition of the model and the analytic solution for the solitonic vortex in an incompressible fluid  followed by the perturbation theory for the effects of compressibility in  the Euler equations. 
Section \ref{sec:vortex} then reports on a number of results for the properties of the solitonic vortex at low orders of perturbation theory. The consequences of these results for the vortex dynamics under the influence of weak harmonic trapping in the quasiparticle dynamics framework are reported in Sec.~\ref{sec:AppEOM} and the relevance of the results for further experimental study are discussed in Sec.~\ref{sec:discussion}. Two appendices discuss the derivation of the Euler equation in the form used in this work and an analytic approximation for the first order correction to the superfluid phase, respectively. 

\section{Hydrodynamic theory} \label{sec:hyd}
\subsection{Solitonic vortex in an incompressible fluid} \label{sec:incompressible}

We consider a three-dimensional uniform slab geometry as sketched in Fig.~\ref{fig:1}a. Hard-wall boundaries confine the superfluid to a rectangular region in the $yz$-plane where the potential is uniform. 
Such a geometry can be realised experimentally with a flat-bottom trap for atomic gases \cite{Gaunt2013}. For simplicity, we assume an infinite extent along the longitudinal $x$-direction \footnote{A weak additional trapping potential in the $x$-direction with a length scale $\gg D$ may be present but is be treated separately in App.~\ref{sec:AppEOM} by means of the local density approximation described in Sec.~\ref{sec:intro}.}, a width $D$ in the $y$-direction and additional confinement $<D$ along the third direction, with which the vortex is aligned. Since the third dimension is irrelevant for the further discussion we will work in the projected two-dimensional $xy$-plane. 
For the geometry of a two-dimensional channel the velocity field $\mathbf{v}_0(x,y)$ of a vortex in an incompressible inviscid fluid can be determined with the method of images  \cite{Saffman1995}. 

The velocity field is related  by $\mathbf{v}_0 = \frac{\hbar}{m}\nabla S_0$ to the superfluid phase $S_0(x,y)$ shown in Fig~\ref{fig:1}b. The phase $S_0$ of the incompressible superfluid with a vortex at position $\boldsymbol{\rho}$ and hard wall boundaries can be constructed by superimposing the known solution for a vortex without boundaries with those of image vortices, such as to satisfy the condition of no flow normal to the boundary in the resulting velocity field. While a single straight wall requires a single image vortex, the two parallel walls of the channel generate a (doubly) infinite array of image vortices. The problem can be solved in an elegant way on the complex plane by introducing the meromorphic velocity potential $w(z)$ with poles as the vortex locations, where $S_0(x,y) = \Re\{w(x+\rmi y)\}$. The  solution for the channel can then be found with the help of a conformal transformation between the channel and the half plane, where a single image vortex suffices \cite{Saffman1995,Greengard1990}:
\begin{equation}
\label{eqn:w}
w = -\rmi q \ln{\left[\frac{\sinh{\left[\frac{\pi}{2D}(z-\rmi h)\right]}}{\sinh{\left[\frac{\pi}{2D}(z-\rmi (2D-h))\right]}}\right]},
\end{equation}
Here, $q \in \field{Z}$ is the quantized charge with $q=1$ for a right-handed singly-charged vortex,
the location of the vortex is $\boldsymbol{\rho} = (0,h)$, and the two-dimensional coordinate space is now represented by the complex plane $z = x + \rmi y$. The $x$-axis is taken parallel to the walls of the channel, the origin is at the bottom wall, and the $y$-axis passes through the vortex. $S_0$ is given by the real part of $w$ and can be written as
\begin{equation} \label{eq:S0}
\tan\left(\frac{S_0 (\textbf{r})}{q}\right) = \frac{\sin{\left(\frac{h\pi}{D}\right)} \sinh{\left(\frac{\pi x}{D} \right)}}{\cos{\left(\frac{\pi y}{D}\right)} - \cos{\left(\frac{h \pi }{D}\right)} \cosh{\left(\frac{\pi x}{D}\right)}}.
\end{equation}
For practical purposes, it is useful to introduce the stream function $\chi(x,y) =  \Im\{w(x+\rmi y)\}$, which has the important property of being single-valued, as opposed to the phase $S_0$. By virtue of the Cauchy-Riemann equations $\partial_x S_0 = \partial_y \chi$ and $\partial_y S_0 = -\partial_x \chi$ any expression involving derivatives of $S_0$ are easily rewritten in terms of $\chi$.

\begin{figure}[t]
\begin{tikzpicture}[baseline]
\pgfmathsetmacro{\xbase}{-4.4}
\pgfmathsetmacro{\ybase}{5.0}
\pgfmathsetmacro{\yoffset}{-2}

\node[scale=0.95] at (0,0) {\includegraphics[width=0.47\textwidth]{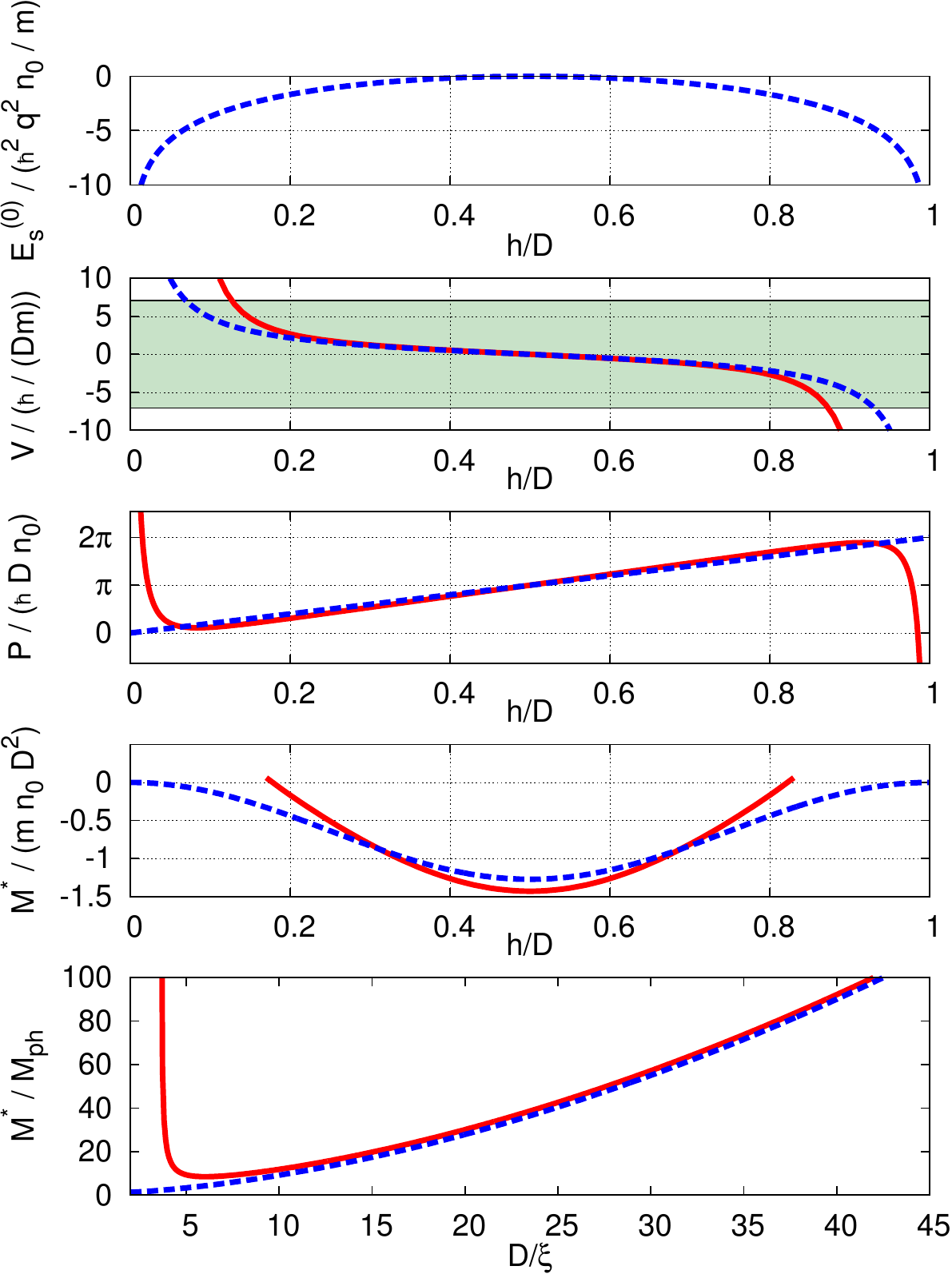}};
\node at (\xbase,\ybase) {(a)};
\node at (\xbase,\ybase+\yoffset) {(b)};
\node at (\xbase,\ybase+2*\yoffset) {(c)};
\node at (\xbase,\ybase+3*\yoffset) {(d)};
\node at (\xbase,\ybase+4*\yoffset) {(e)};

\end{tikzpicture}

\caption{\label{fig:3}
\textbf{Vortex properties.} (\textbf{a}) Renormalized vortex energy $E_\mathrm{s}^{(0)}$ vs. vortex position $h$ from method of images. The vortex velocity (\textbf{b}), canonical momentum (\textbf{c}), and inertial mass (\textbf{d}) are shown in the incompressible limit (dashed blue line) and for the compressible fluid (red full line) with leading order corrections. Here $q=1$ and $\xi^2/D^2 = 10^{-2}$. The shaded region in \textbf{b} shows the sub-sonic regime, where the speed of sound is $c = \pm \frac{1}{\sqrt{2}}\frac{D}{\xi}$. (\textbf{e}) Effective-to-physical mass ratio for the unitary Fermi gas. Shown are the leading-order term $a_2$ (dashed blue line), and the result of including the first-order correction $a_4$ (red full line).}
\end{figure}
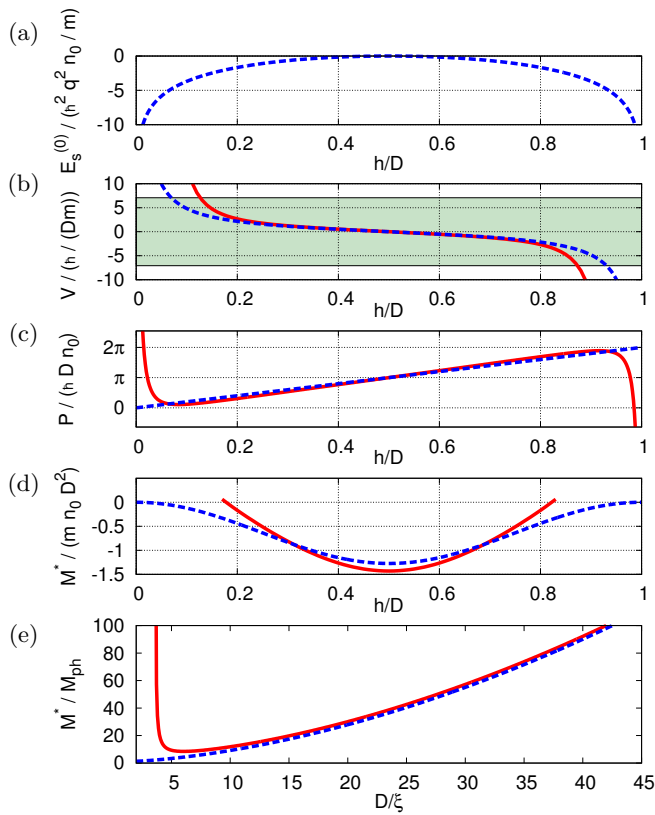

Even though the phase singularity at the position of the vortex results in a locally divergent velocity field, meaning that for example the energy $E_\mathrm{s}$ (Fig.~\ref{fig:3}a) obtained as an integral over the slab of the kinetic energy density is formally divergent, it turns out that our model can be renormalised. Importantly, the compressible corrections to the density and phase fields that we calculate here are renormalisable. This means that we can meaningfully assign several interesting properties of the solitonic vortex; in particular, we find that the vortex moves along the channel in the $x$-direction with a velocity ${V_0} = \frac{\pi q \hbar}{2 Dm}\cot(h\pi/D)$ (Fig.~\ref{fig:3}b), where $q$ is the vortex charge and $h$ the distance from the left channel wall as in  Fig.~\ref{fig:1}a. With the non-divergent canonical momentum $P_0 = 2\pi \hbar q  n_0 h$ (Fig.~\ref{fig:3}c), the inertial mass can be calculated from $M^*_0 = {\d P_0}/{\d {V_0}}$ as 
\begin{equation}
\label{eq:M0}
M^*_0 =  -\frac{4m n_0 D^2}{\pi} \sin^2\left(\frac{\pi h}{D}\right) ,
\end{equation}
where $n_0$ is the two-dimensional density of the incompressible fluid. When the vortex is located at the centre of the channel ($h=D/2$), it is at rest (${V_0}=0$), and the inertial mass $M^*_0 =  -{4m n_0 D^2}/{\pi}$ is simply proportional to the mass of the fluid contained in the square of the channel width $D$.  While the inertia is thus determined by the fluid inside the localisation volume, the negative sign indicates that the solitonic vortex accelerates in the direction opposite to an applied force according to Eq.~\eqref{eq:Newton}.

The incompressible fluid model is directly relevant to real superfluids where it corresponds to the Thomas-Fermi approximation (density follows the local chemical potential) in a flat-bottom trap. Significantly, Eq.~\eqref{eq:M0} for the inertial mass presents an exact result for strongly and weakly-correlated superfluids in the Thomas-Fermi limit $\xi /D \to 0$, in contrast to previous results obtained  for harmonically trapped superfluids \cite{fetter01:vortices,Koens2012,Pitaevskii2013,Ku2014}. A (finite) healing length $\xi$, then, defined by $\d \mu/\d n = \hbar^2/(2nm\xi^2)$, is associated with the compressibility of the fluid. All physical fluids are always compressible.

\subsection{Perturbation expansion for compressible corrections} \label{sec:compressible}

Limitations of the incompressible model appear when we try to evaluate the force term of the equation of motion \eqref{eq:Newton}, wherein the physical mass formally vanishes. A suitable model that overcomes this limitation is that of inviscid isentropic flow described by the Euler equation, which in the co-moving reference frame of the vortex becomes (Appendix~\ref{sec:AppEuler})
\begin{equation}
\label{eq:Euler}
\epsilon \frac{\hbar^2}{2m} |\nabla S |^2 + \mu[n(\textbf{r})] - \mu_0= 0.
\end{equation}
Here $\mu_0$ is the bulk chemical potential, and $\mu[n]$ is the chemical potential in a local density approximation evaluated at $n(\textbf{r})$. For definiteness, we assume a polytropic equation of state 
$n \propto \mu^\gamma$, where $\gamma$ is the polytropic index; for example, for a weakly-interacting Bose-Einstein condensate $\gamma =1$, while for the unitary Fermi gas $\gamma = 3/2$ \cite{Ketterle2008}.
The dimensionless factor  $\epsilon$ is used for collecting orders in a perturbation expansion, and will be set to 1 at the end. 

We note at this point that the Gross-Pitaevskii equation for Bose-Einstein condensates (BECs) can be rewritten as an Euler equation with an additional term known as quantum pressure \cite{P&S} 
\begin{equation} \label{eq:EulerQP}
-\epsilon \frac{\hbar^2}{2m\sqrt{n}}\nabla^2 \sqrt{n} + \epsilon \frac{\hbar^2}{2m} |\nabla S |^2 + \mu[n(\textbf{r})] - \mu_0= 0,
\end{equation}
where we have defined the quantum pressure term (first term on the left) to be of order $\epsilon$. Together with the usual continuity equation (see below), Eq.~\eqref{eq:EulerQP} is fully equivalent to Gross-Pitaevskii theory and provides a quantitative description of dilute-gas BECs. The quantum pressure term governs the detailed structure of solitons and vortex cores. Whether a similar term exists for the crossover superfluid Fermi gas, and the unitary Fermi gas in particular, is not known.

Complemented with the continuity equation
\begin{equation}
\label{eq:continuity}
\nabla \cdot(n \mathbf{v})=0,
\end{equation}
the Euler equation \eqref{eq:Euler} describes a compressible superfluid at zero temperature. Setting $\epsilon=0$ recovers the incompressible problem already solved with the method of images. The more general problem can be solved with a perturbation expansion for the density $n=n_0 + \epsilon n_1 + \epsilon^2 n_2 + \cdots$ and phase $S=S_0 + \epsilon S_1 +\epsilon^2 S_2 + \cdots$. Requiring that $n$ and $S$ solve Eqs.\ \eqref{eq:Euler} and \eqref{eq:continuity}, sorting orders of $\epsilon$ reveals that the density $n_k$ at order $k$ can be expressed through derivatives of the lower-order phase fields, while the phase $S_k$ at order $k$ can be determined from a Poisson equation with a source term containing derivatives of $n_k$ and lower-order phase fields. Through the perturbation expansion, we have essentially linearised the problem by writing the non-linear Euler and continuity equations in a different form as an infinite set of coupled, but linear Poisson equations. By integrating the Poisson equations it is thus possible to determine the phase and density fields to arbitrary order.

Specifically, the density can be obtained using the Euler equation~\eqref{eq:Euler} and inverting the equation of state $\mu[n]$. For definiteness we assume a power-law relation $n_\mathrm{3D} =  \alpha \mu ^\gamma$ for the three-dimensional density~\footnote{Our definition of the polytropic index $\gamma$ differs from the literature by that $\gamma$ is the inverse of the one used in Ref.~\cite{Ketterle2008}.}. Further assuming uniform confinement perpendicular to the $xy$ plane with a length scale $B<D$, we obtain for the two-dimensional density 
\begin{equation}
\label{eqn:Euler123}
n(\textbf{r}) = B \alpha \left[\mu_0 - \epsilon \frac{1}{2}m v^2(\textbf{r})\right]^\gamma.
\end{equation}
The power-law equation of state  covers in particular the case of a weakly-interacting BEC, where $\gamma = 1$ and $\mu = g n_\mathrm{3D}$, where $g$ is the Gross-Pitaevskii coupling so that $\alpha = 1/g$. For the unitary Fermi gas we have $\gamma = 3/2$, and $\alpha = \{{2m}/[{\hbar^2 \xi_{\mathrm{mb}}(3\pi^2)^{\frac{2}{3}}}]\}^{\frac{3}{2}}$, where $\xi_{\mathrm{mb}} \approx 0.37$ is known as the Bertsch parameter \cite{Ku2012}. 
An important length scale related to the equation of state is the healing length $\xi$ with 
$\mu_0 = \gamma \hbar^2 /2 m {\xi^2}$.

We expand $n = n_0 + \epsilon n_1 + \epsilon^2 n_2 + \ldots$, $S = S_0 + \epsilon S_1 + \epsilon^2 S_2 + \ldots$, where $n_k$ and $S_k$ are $k$th order terms in $\epsilon$ in the perturbation expansion. Comparing coefficients of $\epsilon$ in Eq.~\eqref{eqn:Euler123} gives the density corrections
\begin{widetext}
\begin{equation}
\label{eqn:gammancorr}
\begin{split}
n_0 &= B \alpha \mu_0^\gamma,\\
n_1 &=  - \frac{  \hbar^2 }{2 m} \frac{n_0}{\mu_0}  \gamma |\nabla S_0 |^2 ,\\
n_2 &= \frac{ \hbar^2}{8 m}  \frac{n_0}{\mu_0^2} \left[\frac{\gamma !}{(\gamma-2)!}\frac{\hbar^2}{m} |\nabla S_0 |^4 - 8 \frac{\gamma !}{(\gamma-1)!}\mu_0 \nabla S_0 \cdot \nabla S_1\right], \\
n_3 &= \frac{ \hbar^2}{48 m}  \frac{n_0}{\mu_0^3}  \left[-\frac{\gamma !}{(\gamma-3)!} \frac{\hbar^4}{m^2} |\nabla S_0 |^6+24\frac{\gamma !}{(\gamma-2)!} \mu_0 \frac{\hbar^2}{m} |\nabla S_0 |^2 \nabla S_0 \cdot \nabla S_1 -24\frac{\gamma !}{(\gamma-1)!}  \mu_0^2 \left(2 \nabla S_0 \cdot \nabla S_2 +|\nabla S_1|^2\right)\right],\\
n_4 &= \ldots .
\end{split}
\end{equation}
Expressions at arbitrary order can be generated by writing $n$ as a generalised Cauchy product using Mertens' Theorem:
\begin{equation}
\label{eqn:Mertens}
\begin{split}
&n = n_0 \sum_{\ell=0}^\infty \sum_{k_1 = 0}^{\infty} \sum_{k_2 = 0}^{k_1}\ldots \sum_{k_{2\ell} = 0}^{k_{2\ell-1}} {\gamma \choose \ell} \left( \frac{-\hbar^2}{2 m \mu_0}\right)^\ell \epsilon^{\ell} \epsilon^{k_{2\ell}} \epsilon^{k_{2\ell-1}-k_{2\ell}} \nabla S_{k_{2\ell}}\cdot \nabla S_{k_{2\ell-1}-k_{2\ell}} \ldots \epsilon^{k_2-k_3} \epsilon^{k_1-k_2} \nabla S_{k_2-k_3}\cdot \nabla S_{k_1-k_2}.
\end{split}
\end{equation}
\end{widetext}

Similarly, the phase corrections can be deduced using the continuity equation under the condition of stationary flow: $\nabla \cdot (n \textbf{v}) = -\frac{\partial n}{\partial t} = 0$. We get
\begin{equation}
\begin{split}
& [\nabla n_0 + \epsilon \nabla n_1 + \mathcal{O}\left( \epsilon^2 \right)]\cdot [\nabla S_0 + \epsilon \nabla S_1 + \mathcal{O}\left( \epsilon^2 \right)] =\\
& -[n_0 + \epsilon n_1 + \mathcal{O}\left( \epsilon^2 \right)][\nabla^2 S_0 + \epsilon \nabla^2 S_1 + \mathcal{O}\left( \epsilon^2 \right)].
\end{split}
\end{equation}
Clearly $\nabla n_0 = \textbf{0}$, $\nabla^2 S_0 = 0$, and so
\begin{equation}
\label{eqn:Scorr}
\begin{split}
\nabla^2 S_1  &= -\frac{\nabla n_1 \cdot \nabla S_0}{n_0},\\
\nabla^2 S_2  &= -\frac{1}{n_0}\left( \nabla n_1 \cdot \nabla S_1+\nabla n_2 \cdot \nabla S_0 -\frac{n_1}{n_0} \nabla n_1 \cdot \nabla S_0 \right),\\
\nabla^2 S_3 &= \ldots.
\end{split}
\end{equation}

Equations \eqref{eqn:gammancorr} -- \eqref{eqn:Scorr} were derived from the classical Euler equation~\eqref{eq:Euler}. If we include the quantum pressure term for BECs from Eq.~\eqref{eq:EulerQP}, the results are modified by an additive correction to the density expressions of Eq.~\eqref{eqn:gammancorr} at each order in $\epsilon$ with $n_i\to n_i + n_i^{(\mathrm{qp})}$ but otherwise unchanged (Note that $\gamma=1$ for BECs). The quantum pressure corrections can be calculated order by order and read
\begin{equation}
\label{eqn:qpcorr}
\begin{split}
n_0^{(\mathrm{qp})} &= 0,\\
n_1^{(\mathrm{qp})} &= 0 ,\\
n_2^{(\mathrm{qp})} &= - \frac{n_0}{4} \xi^4  \nabla^4   S_0^2,\\
n_3^{(\mathrm{qp})} &= \ldots. \\
\end{split}
\end{equation}

In Fig.~\ref{fig:1}, we show the results for the phase up to first order and the density up to second order. After renormalisation, every order of the phase and all compressible corrections to the density formally diverge only at the vortex position (Appendix~\ref{sec:AppS1}). We note that only $S_0$ has a logarithmic branch point. Even though our counting device $\epsilon$ is not a small parameter itself, it turns out that successive orders in $\epsilon$ also accumulate factors of $\xi^2$, which is small compared to the relevant length scale $D^2$. The addition of a quantum pressure term  to the Euler equation \eqref{eq:Euler}, which would formally make it equivalent to the Gross-Pitaevskii equation (with $\gamma=1$), will affect the density and phase correction terms at order $\mathcal{O}\left(\xi^4/D^4 \right)$.

\section{Solitonic vortex in a compressible superfluid} \label{sec:vortex}
In this section, we obtain several physical properties of the vortex from the compressible hydrodynamics developed in the previous section.

\subsection{\label{sec:vs} Velocity and phase step}
The vortex speed is given by the superfluid flow field at the vortex location after subtracting the free vortex $q/\tilde{r}$ divergence (and any divergences of the velocity corrections),  where $\tilde{r} = |\textbf{r} - \boldsymbol{\rho}|$ is the distance from the vortex. The vortex velocity is parallel to the channel walls:
\begin{equation}
\label{eqn:v}
\begin{split}
\textbf{V} = \frac{\hbar}{m} \lim_{\textbf{r} \to \boldsymbol{\rho}}{\left( \nabla S - \frac{q}{\tilde{r}}\begin{pmatrix} -\sin{(\theta)} \\ \cos{(\theta)} \end{pmatrix} \right)} = V \hat{\mathbf{x}},
\end{split}
\end{equation}
where $\theta$ is an angle measured at the vortex and a series expansion of $V=V_0 +\epsilon V_1 +\cdots$ is defined through the corresponding expansion of the phase $S$. From the incompressible phase $S_0$ of Eq.~\eqref{eq:S0} we easily obtain
\begin{equation}
\label{eqn:v0}
\begin{split}
{V}_0 = \frac{\pi q \hbar}{2D m}\cot{\left( \frac{h \pi}{D} \right)}.
\end{split}
\end{equation}

It turns out that $\nabla S_1$ diverges at the vortex. Expanding the analytical approximation around the vortex,
\begin{equation}
\begin{split}
\nabla S_1^\mathrm{approx} &= -\frac{\pi  q^3 D \hbar }{4m}  \frac{\xi^2}{D^2} \cot \left(\frac{\pi  h}{D}\right) \\
&\times \left( \begin{array}{c}
\frac{\cos (2 \theta )}{\tilde{r}^2} - \frac{\pi ^2 \left[3 \csc ^2\left(\frac{\pi  h}{D}\right)+1\right]}{12 D^2}\\ 
\frac{\sin(2 \theta )}{\tilde{r}^2}  
\end{array}\right)  + \mathcal{O}\left( \tilde{r} \right),
\end{split}
\end{equation}
where we have taken only the leading term for $S_1^\mathrm{approx}$. We can see that once the diverging $\tilde{r}^{-2}$ term is removed, taking the limit $\tilde{r} \to 0$ gives the following first-order correction to the vortex velocity (in the $x$-direction):
\begin{equation}
\label{eqn:v0v1}
\begin{split}
{V}_0 + {V}_1^\mathrm{approx} &=  \frac{\pi q \hbar}{2D m}\cot{\left( \frac{h \pi}{D} \right)} \\
&\times \left\lbrace 1 +  \frac{\pi^2  q^2  }{ 24 }  \frac{\xi^2}{D^2}\left[\frac{3}{\sin ^2\left(\frac{\pi  h}{D}\right)} +1\right] \right\rbrace .
\end{split}
\end{equation}
We show how to obtain the full ${V}_1$ (Fig.~\ref{fig:3}b) numerically below in Sec.~\ref{sec:efmass}.

The phase step across the solitonic vortex can be obtained by evaluating the difference of the limits $(x \to \infty) - (x \to -\infty)$ of the phase field $S_0 +  S_1$. Using the analytical approximation:
\begin{equation}
\label{eqn:phase_step}
\begin{split}
\Delta (S_0 + S_1^\mathrm{approx}) = 2 \pi q \frac{D-h}{D}+ \frac{\pi^2 q^3}{2 } \frac{\xi^2}{D^2}\cot{\left( \frac{h \pi}{D} \right)}.
\end{split}
\end{equation}
A direct numerical evaluation of Eq.\ \eqref{eq:S1int} yields an additional term of order $\xi^2/D^2$ with the same sign.

\subsection{\label{sec:mom}Canonical momentum}

Similar to the case of a dark soliton \cite{pitaevskii03:book} or a vortex ring in a cylindrical wave guide \cite{Pitaevskii2013}, we need to distinguish between the \emph{physical momentum} $P_\mathrm{ph}$  and the \emph{canonical momentum} $P = P_\mathrm{ph} +\hbar n_0 D (2\pi-\Delta S)$. The physical momentum is the integral of the momentum density of the solitonic vortex solution with open boundary conditions for $x\to \pm \infty$ and a phase step $\Delta S$. The canonical momentum is relevant in the context of Hamiltonian dynamics and is obtained by adding the momentum of a back flow current to compensate for the phase step. Indeed the canonical momentum can be obtained directly by integrating the  $x$ component of the velocity field where a background phase gradient has been added such that the solution fulfills periodic boundary condition (pbc) in $x$ direction with box length $L\gg D$: 
\begin{equation}
\begin{split}
P &= m \int_\Omega n v_x^\mathrm{pbc}\,\mathrm{d}^2\textbf{r} \\ \label{eq:Psplit}
& = \hbar \int_\Omega  n_0\partial_x S^\mathrm{pbc}  \,\, \mathrm{d}x\, \mathrm{d}y + \epsilon \hbar \int_\Omega n_1\partial_x S_0^\mathrm{pbc} \,\, \mathrm{d}x\, \mathrm{d}y + \mathcal{O}\left( \epsilon^2 \right)  \\
&= P_0 + \epsilon P_1 + \mathcal{O}\left( \epsilon^2 \right) ,
\end{split}
\end{equation}
where $S^\mathrm{pbc} = S + (2\pi- \Delta S)x/L$.
The first term of the right hand side of Eq.~\eqref{eq:Psplit} is easy to evaluate since $\int \partial_x S_0^\mathrm{pbc} \, \mathrm{d}x$ has the result $2\pi$ for $y<h$ and 0 for $y>h$. The result is (at all orders of $\epsilon$ when expanding $S$) 
\begin{equation}
P_0 = q  h 2 \pi n_0 \hbar .
\end{equation}

There are no branch cuts or logarithmic branch points in $S_1$. In this sense, the vortex is already contained within $S_0$. Hence, the integrations become straightforward as we do not have to worry about crossing any branch cuts. At first order in $\epsilon$ we have
\begin{equation}
\begin{split}
P_1 &= \hbar \int_{\Omega \setminus B_{\boldsymbol{\rho}}^R} n_1 \partial_x S_0^\mathrm{pbc}\, \mathrm{d}^2\textbf{r} \\
&= -\hbar n_0 D^2 \frac{\xi^2}{D^2} \int_{\Omega \setminus B_{\boldsymbol{\rho}}^R}  |\nabla S_0|^2 \partial_x S_0 \,\mathrm{d}^2\textbf{r} \\
&\qquad + \frac{\hbar (2\pi-\Delta S_0)}{L} \int_{\Omega \setminus B_{\boldsymbol{\rho}}^R} n_1 \,\mathrm{d}^2\textbf{r} ,
\end{split}
\end{equation}
where the second term on the right hand side vanishes in the limit $L\to \infty$ and thus can be discarded.
In the region $\Omega \setminus B_{\boldsymbol{\rho}}^R$, we can write the integrand of the first term $I_1$ as a total derivative: 
\begin{equation}
\begin{split}
|\nabla S|^2 \partial_x S &= \nabla \cdot \left[(\partial_x S) \nabla \left(\frac{S^ 2}{2} \right) -\frac{S^2}{2} \nabla (\partial_x S) \right] \\
&=  \nabla \cdot  \left[\chi (\partial_y \chi)  \nabla \chi -\frac{\chi^2}{2}  \nabla (\partial_y \chi) \right].
\end{split}
\end{equation}
The integral is solved by means of the divergence theorem and gives 
\begin{equation}
\begin{split}
P_1 &= \hbar n_0 D^2 \frac{\xi^2}{D^2} \int_{\partial B_{\boldsymbol{\rho}}^R}  \left[\chi (\partial_y \chi) \partial_n \chi -\frac{\chi^2}{2} \partial_n (\partial_y \chi) \right] \,\mathrm{d}s\\
&=  \frac{1}{2}\hbar n_0 D  \pi ^2 q^3     \frac{\xi^2}{D^2} \cot \left(\frac{\pi  h}{D}\right) \left\lbrace1 - 4 \ln \left[\frac{2 D}{\pi  R }\sin \left(\frac{\pi  h}{D}\right)\right]\right\rbrace .
\end{split}
\end{equation}
Finally taking $R = \xi$ we obtain for the momentum
\begin{equation}
\label{eqn:P0}
\begin{split}
P &= 2 q  \hbar  \pi n_0 h + \frac{1}{2}\hbar n_0 D  \pi ^2 q^3     \frac{\xi^2}{D^2} \cot \left(\frac{\pi  h}{D}\right)\\
&\times \left\lbrace1 - 4 \ln \left[\frac{2 D}{\pi  \xi}\sin \left(\frac{\pi  h}{D}\right)\right]\right\rbrace +\mathcal{O}\left(\frac{\xi^4}{D^3}\right).
\end{split}
\end{equation}

\subsection{\label{sec:efmass}Effective mass}
\begin{figure}
\vspace{-0.5cm}
\begin{tikzpicture}[baseline]
\pgfmathsetmacro{\yoffset}{-2.5}
\pgfmathsetmacro{\yoffsetl}{0.13}
\node[scale=0.95] at (-2,0) {\includegraphics[width=0.5\textwidth]{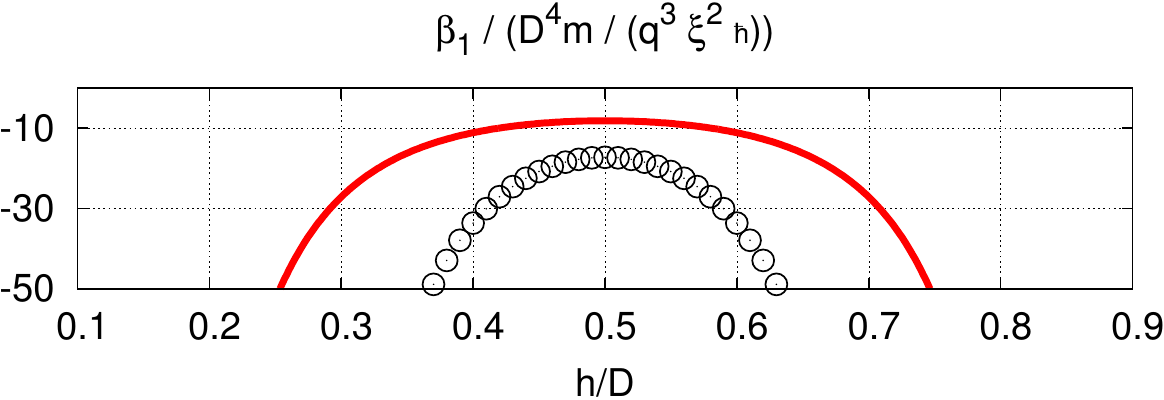}};
\end{tikzpicture}
\caption{\label{fig:beta}
Plot of $\beta_1 = \mathrm{d} V_1 / \mathrm{d} h$, where $V_1$ is the contribution of $S_1$ to $V$, evaluated numerically (black circles) and using $S_1^\mathrm{approx}$ (solid red line). At $h = D/2$, a careful numerical evaluation up to 4 significant figures provides a value consistent with $\tilde{\beta}_1 \approx -(\pi^4/8+\pi^2/2) \approx -17.110$.}
\end{figure}

The effective mass $M^*$ of the vortex can be calculated as
\begin{equation}
\label{eqn:Meffdef}
\begin{split}
M^* &= \frac{\mathrm{d} P}{\mathrm{d} V} = \frac{\mathrm{d} P}{\mathrm{d} h} \left(\frac{\mathrm{d} V}{\mathrm{d} h} \right)^{-1},
\end{split}
\end{equation}
where the canonical momentum $P$ is given in Eq.~\eqref{eqn:P0}. We write the velocity derivative in terms of an expansion $\beta(h) \equiv \mathrm{d} V / \mathrm{d} h = \beta_0 + \beta_1 + \ldots$, where $\beta_i$ corresponds to a term of order $(\xi^2/D^2)^i$. The first term $\beta_0$ is easily found from Eq.~\eqref{eqn:v0} and leads to the inertial mass~\eqref{eq:M0} of the solitonic vortex for the incompressible fluid. Higher order terms can be obtained through Eqs.~\eqref{eqn:v} by integration. To order $\xi^2/D^2$, we obtain $\beta_1 = \tilde{\beta}_1 {q^3 \xi^2 \hbar}/{D^4 m}$, where $\tilde{\beta}_1\left( h/D\right)$ is a dimensionless function shown in Fig.~\ref{fig:beta}. At $h = D/2$, we obtain numerically $\tilde{\beta}_1\left( \frac{1}{2}\right) = -17.1109\approx -\left(\frac{\pi^4}{8} +\frac{\pi^2}{2}\right)$.

Evaluating Eq.~\eqref{eqn:Meffdef}, we obtain
\begin{equation}
\begin{split}
M^* &= -\frac{4}{\pi}  D^2 m n_0 \sin ^2\left(\frac{\pi  h}{D}\right)\\
& + m  \pi n_0 q^2 \xi ^2 \left\lbrace -4 \ln \left[\frac{2 D}{\pi  \xi  }\sin \left(\frac{\pi  h}{D}\right)\right]+2 \cos \left(\frac{2 \pi  h}{D}\right) \right.\\
&\qquad \left. +3 - \frac{8\tilde{\beta}_1}{\pi^4}  \sin ^4\left(\frac{\pi  h}{D}\right)\right\rbrace +\mathcal{O}\left(\frac{\xi^4}{D^2}\right).
\end{split}
\end{equation}
For the effective mass of a stationary solitonic vortex at $h = D/2$, this yields
\begin{equation}
\label{eqn:Meff_1storder_numerical}
\begin{split}
\left.M^*\right|_{h= \frac{D}{2}} &=  D^2 m n_0 \left\{  -\frac{4}{\pi} -   q^2 \frac{\xi^2}{D^2} \left[ 4\pi \ln \left(\frac{2 D}{\pi  \xi  }\right) \right. \right.\\
&\left. \left. - 2\pi -  \frac{4}{\pi} + \Xi_1\right] +\mathcal{O}\left(\frac{\xi^4}{D^2}\right) \right\},
\end{split}
\end{equation}
where $\left| \Xi_1  \right| \lesssim 10^{-4}$, as determined by our numerical accuracy. We note that the first-order correction to $\left.M^*\right|_{h= \frac{D}{2}}$ has the same sign as the leading-order term provided that $\frac{\xi}{D} < \frac{2 }{\pi }\rme^{-\frac{2+\pi ^2}{2 \pi ^2}} \approx 0.349$, making the effective mass even more negative.

By definition, $\beta(h) = \mathrm{d} V / \mathrm{d}h$. For $\beta_1$, we need $\textbf{V}_1 = (\hbar/m) \nabla S_1$ at the vortex with the divergent `free vortex' terms removed as in Appendix~\ref{sec:AppS1} and Sec.~\ref{sec:vs}. Using Eq.~\eqref{eqn:S1_int}, we start by writing
\begin{equation}
\begin{split}
\beta_1 &= \frac{\hbar}{m} \frac{\mathrm{d} (\partial_x S_1)}{\mathrm{d}h}  \\
&= \frac{\hbar}{m} \frac{\mathrm{d}}{\mathrm{d}h} \int_{\Omega}  \partial_x \left[ -\frac{1}{n_0} \nabla \cdot (n_1(\textbf{r}^\prime) \nabla S_0(\textbf{r}^\prime)) G(\textbf{r},\textbf{r}^\prime)  \right] \mathrm{d}^2\textbf{r}^\prime.
\end{split}
\end{equation}
As before for $S_1$, we renormalise $\beta_1$ by expanding the integrand about the vortex position as a power series in the distance $\tilde{r}$ from the vortex. The diverging terms are analytic expressions that scale with inverse powers of the distance $\tilde{r}$, from $\tilde{r}^{-4}$ to $ \tilde{r}^{-1}$. We then renormalise the original integrand by adding four counter terms. We integrate the resulting finite integrand numerically to obtain $\textbf{V}_1$, which we then differentiate with respect to $h$ to get $\beta_1$. The numerically thus evaluated $\textbf{V}_1$ (whose only component is along $x$) is shown in Fig.~\ref{fig:3}b as the red solid line.

\subsection{Number of missing particles and physical mass} \label{sec:physicalmass}
Vortex lines are usually associated with a core region of depleted particle number density due to large kinetic energy densities near the vortex filament. This depletion is quantified by the number of missing particles 
\begin{equation} \label{eq:Ns}
N_{\mathrm{s}} = \int_{\Omega } (n-n_0) \,\mathrm{d}^2\textbf{r} ,
\end{equation}
which usually takes negative values. The missing particle number  $N_{\mathrm{s}}$ is closely related to the physical mass defined by Eq.~\eqref{eqn:physMassdef}. In fact, for a zero-velocity solitonic vortex \cite{Scott2011,Schecter2012,Shamailov2016}
\begin{equation} \label{eq:mph}
M_{\mathrm{ph}} = m N_{\mathrm{s}} ,
\end{equation}
where the zero-velocity condition is equivalent to setting $h=D/2$. In the following we outline the procedure for calculating  $N_{\mathrm{s}}$.

Since the individual terms $n_i$ and any finite approximation to the series expansion of the density  diverge near the vortex position, the integral \eqref{eq:Ns} has to be renormalized. As a consequence, the approximate density may become negative in a small region near the vortex position. A renormalisation procedure can then be defined by excluding the area $B$ from the integral where the approximate density has negative values, defined by $n(\mathbf{r})=0$ at $\mathbf{r}\in \partial B$. With $n=n_0 +n_1$ and  $\xi/D \to 0$ this condition yields a  a disk  $B_{\boldsymbol{\rho}}^R$ with a radius $R=\xi$. Assuming convergence of  the infinite series expansion for the density $n$, the excision radius $R$ will shrink to zero, as the density in the hydrodynamic description of a compressible superfluid vanishes only at the position of the vortex.

We start by formally expanding $N_{\mathrm{s}}$ in powers of $\epsilon$:
\begin{equation}
\begin{split}
N_{\mathrm{s}} &= \int_{\Omega \setminus B_{\boldsymbol{\rho}}^R} (n-n_0)\,\mathrm{d}^2\textbf{r} \\
&= \int_{\Omega \setminus B_{\boldsymbol{\rho}}^R}\left[\epsilon n_1 + \epsilon^2 n_2 + \mathcal{O}\left( \epsilon^3 \right)\right] \,\mathrm{d}^2\textbf{r} \\
&=  N_{\mathrm{s}}^{(\epsilon)} + N_{\mathrm{s}}^{(\epsilon^2)} + \mathcal{O}\left( \epsilon^3 \right).
\end{split}
\end{equation}
From Eq.~\eqref{eqn:gammancorr}, considering only $n_1$, we get
\begin{equation}
\begin{split}
\label{eqn:N_s_direct}
N_{\mathrm{s}}^{(\epsilon)} &= -n_0 D^2\frac{\xi^2}{D^2}  \int_{\Omega \setminus B_{\boldsymbol{\rho}}^R} |\nabla S_0|^2 \,\mathrm{d}^2\textbf{r} \\
&= -2 \pi q^2 n_0 \xi^2 \ln{\left[\frac{D}{R}\sin{\left(\pi \frac{h}{D} \right)} \right]}.
\end{split}
\end{equation}
Considering $n_2$, we can write
\begin{equation}
\begin{split}
N_{\mathrm{s}}^{(\epsilon^2)} &= - 2n_0\xi^2 \int_{\Omega \setminus B_{\boldsymbol{\rho}}^R} \nabla S_0 \cdot \nabla S_1 \,\mathrm{d}^2\textbf{r} \\
&\qquad + \frac{\gamma - 1}{\gamma} \frac{n_0 \xi^4}{2} \int_{\Omega \setminus B_{\boldsymbol{\rho}}^R} |\nabla S_0|^4\,\mathrm{d}^2\textbf{r}, \\
\end{split}
\end{equation}
but this integral is more difficult to evaluate in closed form, and essentially would give $S_1$ as well. 

The procedure of excising a disk of radius $R$, which we then identify with the healing length $\xi$, introduces complications with terms of order $\epsilon^2$ and higher. In particular, at $h = D/2$, we find both analytically and numerically the following dependence on $R$:
\begin{equation}
\label{eqn:ps}
\begin{split}
\int_{\Omega \setminus B_{\boldsymbol{\rho}}^R} |\nabla S_0|^4 \,\mathrm{d}^2\textbf{r}&= \frac{\pi q^4}{R^2} - \frac{\pi^5 q^4 }{9 D^4} R^2 + \mathcal{O}\left( \frac{R^4}{D^6}\right),\\
\int_{\Omega \setminus B_{\boldsymbol{\rho}}^R} |\nabla S_0|^6\,\mathrm{d}^2\textbf{r} &= \frac{\pi  q^6}{2 R^4}-\frac{\pi ^5 q^6}{2 D^4}\ln \left(\frac{R}{D}\right) + \mathcal{O}\left( \frac{R^4}{D^6}\right), \\
\int_{\Omega \setminus B_{\boldsymbol{\rho}}^R} |\nabla S_0|^n\,\mathrm{d}^2\textbf{r} &= \frac{2\pi  q^n}{(n-2) R^{n-2}} + \ldots,\quad n = 4, 6, \ldots .
\end{split}
\end{equation}
The leading order terms do not depend on $h$. After the identification $R = \xi$, this means that we need all the terms in the perturbation series just to get the leading-order $\mathcal{O}(\xi^2)$ term for $N_{\mathrm{s}}$. Fortunately, it is possible to evaluate this contribution from all the terms of order $\epsilon^2$ and higher using the general expression~\eqref{eqn:Mertens}. Summing over all the higher-order contributions of integrals of the form $\int_{\Omega \setminus B_{\boldsymbol{\rho}}^R} |\nabla S_0|^n\,\mathrm{d}^2\textbf{r}$, we obtain
\begin{equation}
\begin{split}
\label{eqn:NsnumUFG}
N_\mathrm{s} &= -2 \pi n_0  q^2 \xi^2 \ln{\left[\frac{D}{\xi}\sin{\left(\pi \frac{h}{D} \right)} \right]} \\
&+\, _pF_q\left(1,1,2-\gamma;2,3; \frac{1}{\gamma}\right) \frac{\gamma - 1}{2\gamma} n_0 \pi q^2 \xi^2 \\
&- 2n_0\xi^2 \int_{\Omega \setminus B_{\boldsymbol{\rho}}^R} \nabla S_0 \cdot \nabla S_1 \,\mathrm{d}^2\textbf{r} + \mathcal{O}\left( \frac{\xi^6}{D^4}\right),
\end{split}
\end{equation}
where $_pF_q$ is the generalised hypergeometric function. With the unitary Fermi gas ($\gamma = 3/2$), we have $_pF_q\left(1,1,\frac{1}{2};2,3; \frac{2}{3}\right) \approx 1.06829$. For the case of a BEC ($\gamma = 1$), the second term in Eq.~\eqref{eqn:NsnumUFG} drops out, and the quantum pressure terms only contribute at order $\mathcal{O}\left( \frac{\xi^4}{D^2}\right)$. At $h = D/2$, numerical evaluation shows that
\begin{equation}
\label{eqn:Nseps2}
\int_{\Omega \setminus B_{\boldsymbol{\rho}}^R} \nabla S_0 \cdot \nabla S_1 \,\mathrm{d}^2\textbf{r} = -2\pi  \Xi_2 q^4  \frac{\xi ^2}{D^2} \ln{\left( \frac{D}{R}\right)} + \ldots,
\end{equation}
where $\left| \Xi_2 - 5 \right| \lesssim 10^{-2}$, as determined by our numerical accuracy.

The quantum pressure corrections contribute to $n_2$, and so have an effect on the missing particle number at order $\epsilon^2$ and higher. As is the case without the quantum pressure, the excision procedure introduces problems when evaluating $N_\mathrm{s}$. For example, we can analytically evaluate the first quantum pressure correction
\begin{equation}
\begin{split}
\int_{\Omega \setminus B_{\boldsymbol{\rho}}^R}n_2^{(\mathrm{qp})}\,\mathrm{d}^2\textbf{r} &= - \frac{n_0}{4} \xi^4  \int_{\Omega \setminus B_{\boldsymbol{\rho}}^R}\nabla^4   S_0^2 \,\mathrm{d}^2\textbf{r} \\
&\stackrel{R = \xi}{=} -2 \pi  n_0q^2 \xi^2 + \mathcal{O}\left(\frac{\xi^6}{D^4} \right),
\end{split}
\end{equation}
which, as before, means that we must evaluate higher-order corrections to obtain the full contribution to $N_{\mathrm{s}}^{(\epsilon)}$. While this is possible in principle, we have not performed a fully detailed calculation to obtain this contribution.

\section{\label{sec:AppEOM} Equation of motion under harmonic trapping}

In the previous Secs.~\ref{sec:hyd} and \ref{sec:vortex} we have considered a homogeneous slab extending infinitely in the $x$-direction, which led to the vortex moving at constant velocity parallel to the $x$-axis. In this section we consider the effects of a weak harmonic trapping potential  $V_\mathrm{trap}(x) = \frac{1}{2} m_p \omega_x^2 x^2$, where $m_p$ is the mass of the elementary particles confined by the trapping potential \footnote{In the case of fermionic superfluids $2m_p=m$, where $m$ is the mass of a Cooper pair. For bosons $m_p=m$.}. An important property of the solitonic vortex is that it is localised along the $x$-axis. Indeed it can be seen from Eq.~\eqref{eq:S0} that the incompressible phase field $S_0(\mathbf{r})$ approaches a constant value (a vacuum) exponentially with a characteristic length scale $D/\pi$ on either side of the vortex core and the exponential localisation remains valid at all orders of the perturbation expansions \eqref{eqn:gammancorr} and \eqref{eqn:Scorr} for the density and phase of the compressible superfluid solutions. For this reason, in the presence of a trapping potential, the dynamics of the solitonic vortex can be treated in the framework of Landau quasiparticle dynamics, as outlined in Sec.~\ref{sec:intro} and previously considered in Ref.~\cite{Ku2014}.

\subsection{Period of small amplitude oscillations}

\begin{figure}[t]
\begin{tikzpicture}[baseline]

\node at (-5,1.3) {(a)};
\node[scale=0.95] at (-3,-1) {\includegraphics[width=0.25\textwidth]{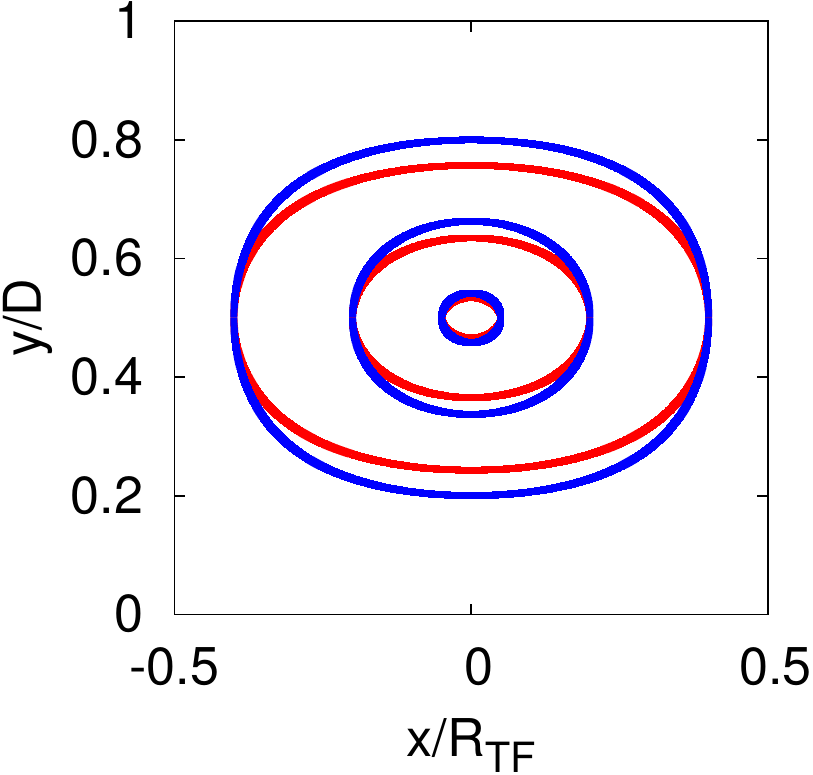}};
\node[scale=0.95] at (1.5,-1) {\includegraphics[width=0.25\textwidth]{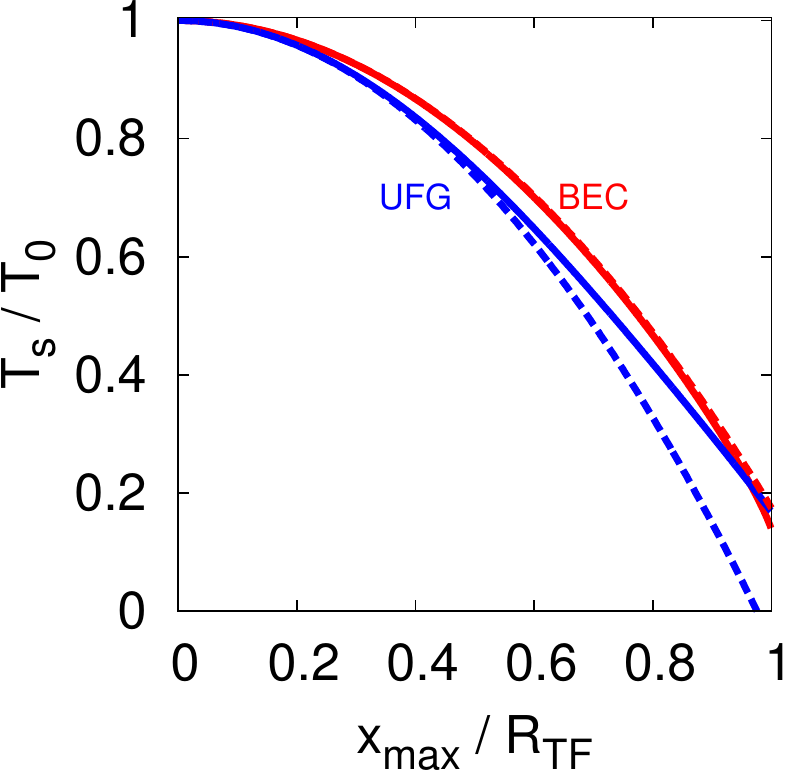}};
\node at (-0.5,1.3) {(b)};

\end{tikzpicture}
\caption{\label{fig:2}
\textbf{Vortex oscillations.} Typical vortex trajectories (\textbf{a}) and oscillation periods as a function of amplitude (\textbf{b}) for a Bose-Einstein condensate (BEC) (red lines) and a unitary Fermi gas (UFG, blue lines) under the influence of a longitudinal trapping potential $V_\mathrm{trap}=\frac{1}{2}m \omega_\mathrm{x}^2 x^2$ with $\xi/D = 0.1$. The trajectories are obtained from a local density approximation  and conservation of energy. The Thomas-Fermi radius is given by $R_{\mathrm{TF}}^2=2\mu_0/(m\omega_\mathrm{x}^2)$. The oscillation period is evaluated with the leading-order (incompressible) vortex velocity, and is shown relative to the period of small-amplitude oscillations $T_0$, which is given by the mass ratio at $h = D/2$. The period $T_\mathrm{s}$ depends on the oscillation amplitude, but for amplitudes below $0.1 R_{\mathrm{TF}}$ the motion is approximately simple harmonic. The dashed lines in (\textbf{b}) show the analytic expansion $\frac{T_\mathrm{s}}{T_0} = 1 + \frac{\gamma -2 \gamma  \ln\left(\frac{D}{\xi }\right)-3}{8} \left(\frac{x_\mathrm{max}}{R_\mathrm{TF}}\right)^2  + \mathcal{O}\left(\frac{x_\mathrm{max}}{R_\mathrm{TF}}\right)^4$.}
\end{figure}

In the harmonic trapping potential the buoyancy-like restoring force $F$ takes the form 
\begin{align}
F= -\frac{m}{m_p} M_\mathrm{ph} \omega_x^2 x ,
\end{align}
and the solutions of Newton's equation \eqref{eq:Newton} are oscillations, as shown in Fig.~\ref{fig:2}. In the regime of small-amplitude oscillations, the inertial and physical masses $M^*$ and $M_\mathrm{ph}$ can be replaced by their respective values in the center of the channel. Equation  \eqref{eq:Newton} now becomes that of a harmonic oscillator with an oscillation period $T_{0}$ given by
\begin{equation}
\label{eqn:massratio}
\begin{split}
\left( \frac{T_{0}}{T_{\mathrm{trap}}} \right)^{-2} &= \left.\frac{M_\mathrm{ph}}{ M^*}\right|_{h=\frac{D}{2}} \\
&= {a_2 \frac{\pi ^2q^2}{2 } \frac{\xi^2}{D^2} + a_4\frac{\pi^2 q^4}{2}\frac{\xi^4}{ D^4}  +\mathcal{O}\left( \frac{\xi^6}{D^6} \right) }, 
\end{split}
\end{equation}
where $T_{\mathrm{trap}} = 2\pi/\omega_\mathrm{x}$. The coefficients $a_2$ and $a_4$ are at most logarithmic functions of  $\xi^2/D^2$ with explicit expressions as follows: 
\begin{equation}
\begin{split}
a_2 &= \ln{\left(\frac{D}{\xi} \right)} - F,\\
a_4 &= -\left[  F + 9\ln \left(\frac{D}{\xi }\right)\right] \\
&\qquad + \left[\ln \left(\frac{D}{\xi }\right) - F\right] \left[ \pi ^2 \ln \left(\frac{\pi  \xi }{2 D}\right)+ \frac{\pi ^2}{2} \right],
\end{split}
\end{equation}
where $F \equiv \, _pF_q\left(1,1,2-\gamma ;2,3;\frac{1}{\gamma }\right)  \frac{\gamma-1}{4\gamma}$.
For the unitary Fermi gas $\gamma=\frac{3}{2}$ and $F\approx 0.08902$. In the BEC case ($\gamma = 1$) $F=0$ and quantum pressure terms are required to obtain the full coefficients. In principle, the coefficients can be evaluated to any order.

\subsection{Trajectories as energy contours}

Beyond the limit of small-amplitude oscillations, the oscillation period of the solitonic vortex depends on the amplitude. The trajectories of the solitonic vortex and its oscillation period can be calculated from the free energy $E_\mathrm{s}(\mu, V)$ in local density approximation (LDA) where $\mu$ is replaced by the local chemical potential at the position of the vortex  $\mu(X) = \mu_0- V_\mathrm{trap}(X)$. Evaluating the energy expression in the homogeneous slab
\begin{align}
  E_\mathrm{s} = \frac{\hbar^2 }{2m} \int_{\Omega \setminus B_{\boldsymbol{\rho}}^\xi} n|\nabla S|^2 \,\mathrm{d}^2\textbf{r},
\end{align}
to leading order yields
\begin{equation}
\label{eqn:F0}
\begin{split}
E_\mathrm{s}^{(0)} &= \frac{\hbar^2 }{2m}n_0 \int_{\Omega \setminus B_{\boldsymbol{\rho}}^R} |\nabla S_0|^2 \,\mathrm{d}^2\textbf{r} \\
&= \frac{\hbar^2 q^2}{m}\pi n_0 \ln{\left[\frac{D}{R} \sin{\left(\pi \frac{h}{D} \right)} \right]},
\end{split}
\end{equation}
where we have regularised the integral by excising a disk with radius $R$ as in App.~\ref{sec:AppS1}.
In the presence of the harmonic trapping potential by virtue of the LDA the background density $n_0$ is replaced by
\begin{equation}
n_{\mathrm{LDA}}(x) = n_0 \left(1 - \frac{x^2}{R_{\mathrm{TF}}^2} \right)^\gamma,
\end{equation}
where we have made use of the polytropic equation of state, and $R_{\mathrm{TF}}=\sqrt{2\mu_0/(m\omega_\mathrm{x}^2)}$ is the Thomas-Fermi radius. Applying the LDA to the energy expression of Eq.~\eqref{eqn:F0} and identifying the cutoff radius $R$ with the healing length $\xi$, we obtain
\begin{equation}
\label{eqn:trajectory}
E_\mathrm{s,LDA}^{(0)}(x,y) = \frac{\hbar^2 q^2}{m}\pi n_0 \left(1 -\frac{x^2}{R_{\mathrm{TF}}^2}\right)^\gamma\ln{\left[\frac{D}{\xi} \sin{\left(\pi \frac{y}{D} \right)} \right]},
\end{equation}
where $x$ and $y\equiv h$ are the coordinates of the vortex position in the slab.
Vortex trajectories are given by $E_\mathrm{s,LDA}^{(0)} = f_0$, where $f_0$ is a positive constant. The higher the energy, the slower the vortex as its effective mass is negative, and the smaller the amplitude of the trajectory.

Instead of the constant $f_0$, we can use the turning points to have $x_\mathrm{max}$, the amplitude of the oscillations, as our control parameter. Eliminating $f_0$ using Eq.~\eqref{eqn:trajectory}, we obtain an implicit function for $y$ and $x$ in terms of $x_\mathrm{max}$ that gives the same trajectories:
\begin{equation}
\label{eqn:trajectory_xmax}
\sin{\left(\pi \frac{y}{D} \right)} = \frac{\xi}{D} \mathrm{exp} \left[\ln{\left(\frac{D}{\xi}\right) }\left(\frac{1 -x_\mathrm{max}^2/R_{\mathrm{TF}}^2 }{1 -x^2/R_{\mathrm{TF}}^2} \right)^\gamma \right].
\end{equation}

\subsection{Oscillation period}
The oscillation period can be calculated by integration over the trajectory as
\begin{align}
 T_\mathrm{s} = 4  \int_0^{x_\mathrm{max}} \frac{\mathrm{d}x}{V}.
\end{align}
Figure~\ref{fig:2}(b) shows the resulting dependence of the oscillation period on the amplitude $\mathrm{max}$ obtained from numerical integration, where the velocity $V=V_0 + V_1$ includes numerically obtained compressible  corrections up to $\mathcal{O}(\xi^2/D^2)$. 

An analytic approximation for the oscillation period can be obtained by approximating the velocity by $V = V_0$ as given in Eq.~\eqref{eqn:v0}.  Integrating over the energy contour \eqref{eqn:trajectory_xmax} and taking the limit $x_\mathrm{max}\to 0$ reproduces $T_0$ of Eq.~\eqref{eqn:massratio} to leading order in $\xi^2/D^2$. 
Expanding around $x_\mathrm{max} = 0$, we obtain
\begin{widetext}
\begin{equation}
\frac{T_\mathrm{s}}{T_0} = 1 + \frac{\gamma -2 \gamma  \ln\left(\frac{D}{\xi }\right)-3}{8} \left(\frac{x_\mathrm{max}}{R_\mathrm{TF}}\right)^2  + \frac{\gamma  (\gamma +10)+4 \gamma  \ln \left(\frac{D}{\xi }\right) \left(3 \gamma +\gamma  \ln \left(\frac{D}{\xi }\right)-5\right)-23}{256 }\left(\frac{x_\mathrm{max}}{R_\mathrm{TF}}\right)^4  + \cdots .
\end{equation}
\end{widetext}
The second order expansion of the oscillation period is shown in Fig.~\ref{fig:2}(b) as dashed lines.

\section{Discussion} \label{sec:discussion}

The mass ratio~\eqref{eqn:massratio} can become large for a solitonic vortex compared to a dark soliton, which is consistent with the interpretation of recent experiments~\cite{Ku2014,Serafini2015} performed in cylindrical, all-harmonic trapping potentials. Future experiments in a rectangular slab geometry with hard-wall potentials could provide quantitative comparison of the oscillation period with the results of this work -- potentially at very high accuracy. This could lead to identifying effects that are not yet included in the theory including  the Kopnin mass, a mass contribution to the vortex expected to arise from fermionic bound states in the vortex core of a fermionic superfluid \cite{Sonin2013}. Also the role of quantum pressure contributions to the hydrodynamic equations of Fermi gases in the BEC to BCS crossover is currently unclear and experiments or Monte-Carlo simulations could clarify the situation by comparison to the theory presented here. 

Our results for the inertial vortex mass may be compared with the recent discussion of vortex mass contributions by Sonin \cite{Sonin2013}. Based on physical arguments but without recourse to a specific confining geometry, Sonin introduces the `core' mass and the `compressibility' mass.
While no mass contribution that scales as the leading contribution to the inertial mass \eqref{eqn:Meff_1storder_numerical} appears in Ref.~\cite{Sonin2013}, the `core' and `compressibility' masses resemble the  next to leading order contributions $\propto m q^2 n_0 \xi^2$ to the inertial mass  of Eq.~\eqref{eqn:Meff_1storder_numerical}. The `core' mass is related to the fluid displacement by the vortex core and is consistent with the non-logarithmic contribution to the $\xi^2$ term in Eq.~\eqref{eqn:Meff_1storder_numerical}. It is not possible to compare the terms quantitatively as any constant can be absorbed in the logarithmic term that depends on the geometry. Sonin's `compressibility' mass has the same scaling as but a different sign than the $\xi^2 \ln(2D/\pi\xi)$ term in Eq.~\eqref{eqn:Meff_1storder_numerical}, which does depend on the confining geometry. We find that the overall sign of this term becomes negative when $\xi/D \lesssim 0.35$ whereas both contributions discussed in Ref.~\cite{Sonin2013} as being relevant for bosonic superfluids are positive. It should be noted that the contributions at this level are geometry-dependent and thus it is essential to define the geometry as done with the slab geometry in this paper in order to make quantitative predictions.

The series expansion for the physical and inertial vortex masses established here provides the basis to further improving the theoretical understanding of strongly-correlated quantum liquids like the superfluid Fermi gas beyond the currently known hydrodynamic model \eqref{eq:Euler}. By measuring the oscillation period $T_0$, future high-precision experiments in the slab geometry can determine the expansion coefficients order-by-order and inform the modelling of quantum pressure, transverse forces \cite{Iordansky1964a,Thouless1996,Kopnin2002}, and vortex core filling~\cite{LAT_JB_15_1}. Moreover, our results can be further developed to make experimental measurements of certain microscopic features such as the level spacing of the Andreev bound states of the vortex core~\cite{LAT_JB_15_1}. 

\textbf{Acknowledgements}: The authors thank Sandy Fetter for inspiring discussions. Both of the authors contributed equally to the conception of the work and to the writing of the manuscript. L.A.T. carried out the detailed calculations and J.B. designed the incompressible model.

\appendix
\section{\label{sec:AppEuler} Derivation of the Euler equation}
Here we outline the derivation of Eq.~\eqref{eq:Euler} in the main text. At zero temperature, changes in the chemical potential $\mu$ for a bulk system are related to changes in the pressure $p$ by the Gibbs-Duhem relation $\d p =  n \d \mu / m $, where $n$ is the number density. The third law of thermodynamics states $s = 0$ for a perfect crystal at $T = 0$. On the other hand, for an isentropic process ($\d s = 0$, where $s$ is the entropy), we have $\d p = n \d \tilde{w}$~\cite{landau1987}, where $\tilde{w}$ is the enthalpy. Let us therefore identify $\mu = m \tilde{w}$ at zero temperature for isentropic processes. The Nernst-Simon formulation of the third law of thermodynamics states that $\d s \to 0$ for any reversible isothermal ($\d T = 0$) process as $T \to 0$, i.e. any reversible isothermal process at $T = 0$ is isentropic. 

The Euler equation in Eq. (2.9) of Ref.~\cite{landau1987} is given as:
\begin{equation}
\frac{\partial \textbf{v}}{\partial t} + \nabla \tilde{w} + (\textbf{v} \cdot \nabla) \textbf{v} = \textbf{0}.
\end{equation}
Applying a Galilean boost $\textbf{r} = \textbf{r}^\prime + \textbf{v}_\mathrm{b} t$, $t = t^\prime$, where $\textbf{v}_\mathrm{b}$ is a constant Galilean boost velocity, i.e. $\textbf{v} = \textbf{v}^\prime + \textbf{v}_\mathrm{b}$, we get $\frac{\partial}{\partial t^\prime} = \textbf{v}_\mathrm{b} \cdot \nabla + \frac{\partial}{\partial t}$, and $\nabla = \nabla^\prime$, and so
\begin{equation}
\frac{\partial \textbf{v}^\prime}{\partial t^\prime} - (\textbf{v}_\mathrm{b} \cdot \nabla^\prime )\textbf{v}^\prime + \nabla^\prime \tilde{w} + [(\textbf{v}^\prime + \textbf{v}_\mathrm{b}) \cdot \nabla^\prime] \textbf{v}^\prime = \textbf{0},
\end{equation}
where we have used $\dot{ \textbf{v}}_\mathrm{b} = \textbf{0}$ and that its gradients also vanish. Importantly, in the boosted frame that travels alongside the vortex, the velocity does not change, i.e. $\frac{\partial \textbf{v}^\prime}{\partial t^\prime} = \textbf{0}$. In the main text we calculate $\textbf{v}^\prime$. We obtain (dropping the primes)
\begin{equation}
\nabla \tilde{w} + (\textbf{v}\cdot \nabla) \textbf{v} = \textbf{0}.
\end{equation}
Using $\frac{1}{2}\nabla (\textbf{v} \cdot \textbf{v}) =  (\textbf{v}\cdot \nabla) \textbf{v} + \textbf{v} \times (\nabla \times \textbf{v}) = (\textbf{v}\cdot \nabla) \textbf{v} $ where $\nabla \times \textbf{v} = \nabla \times (\nabla S) = \textbf{0}$, we can rewrite the above equation as follows:
\begin{equation}
\label{eqn:Be}
\nabla \tilde{w} + \frac{1}{2}\nabla v^2 = \textbf{0}.
\end{equation}
This is nothing more but the Bernoulli equation for a compressible fluid, Eq. (5.3) in Ref.~\cite{landau1987}:
\begin{equation}
\label{eqn:Be1}
\frac{1}{2}v^2 + \tilde{w} = \mathrm{const}.
\end{equation}
Equation~\eqref{eq:Euler} of the main text now follows after multiplying with m by replacing $m \tilde{w}$ with the chemical potential at the local density and denoting the constant with $\mu_0$.

\section{\label{sec:AppS1} First-order compressible phase correction $S_1$: regularisation and renormalisation}
\subsection{Poisson equation for  $S_1$}
The first-order velocity correction can be obtained from Eq.~\eqref{eqn:Scorr}, which amounts to solving Poisson's equation in the domain $\Omega$ with zero Neumann boundary conditions (where $\Omega$ represents the channel as per Fig.\ 1a of the main text):
\begin{equation}
\label{eqn:s111}
\begin{split}
\nabla^2 S_1 &= -\frac{1}{n_0} \nabla n_1\cdot \nabla S_0 \\
&= \frac{\hbar^2}{2m}\frac{\gamma}{\mu_0} \nabla \cdot (|\nabla S_0 |^2 \nabla S_0) + \frac{\hbar^2}{2m}\frac{\gamma}{\mu_0} |\nabla S_0 |^2  \nabla^2 S_0, \\
\end{split}
\end{equation}
where $S_1$ is the first order phase field, and $n_1 = -\frac{\hbar^2}{2m}\frac{n_0}{\mu_0} \gamma|\nabla S_0|^2$ is the first-order density correction. 
Note that the right hand side would be zero for a free vortex since there the velocity field lines $\nabla S_0$ are orthogonal to the density gradient $\nabla n_1$. For the solitonic vortex this is not the case and thus the right hand side is non-zero.

\begin{figure*}
\vspace{-1.5cm}
\begin{tikzpicture}[baseline]
\pgfmathsetmacro{\yoffset}{-2.5}
\pgfmathsetmacro{\yoffsetl}{0.13}
\node[scale=0.95] at (0,\yoffset) {\includegraphics[width=0.47\textwidth]{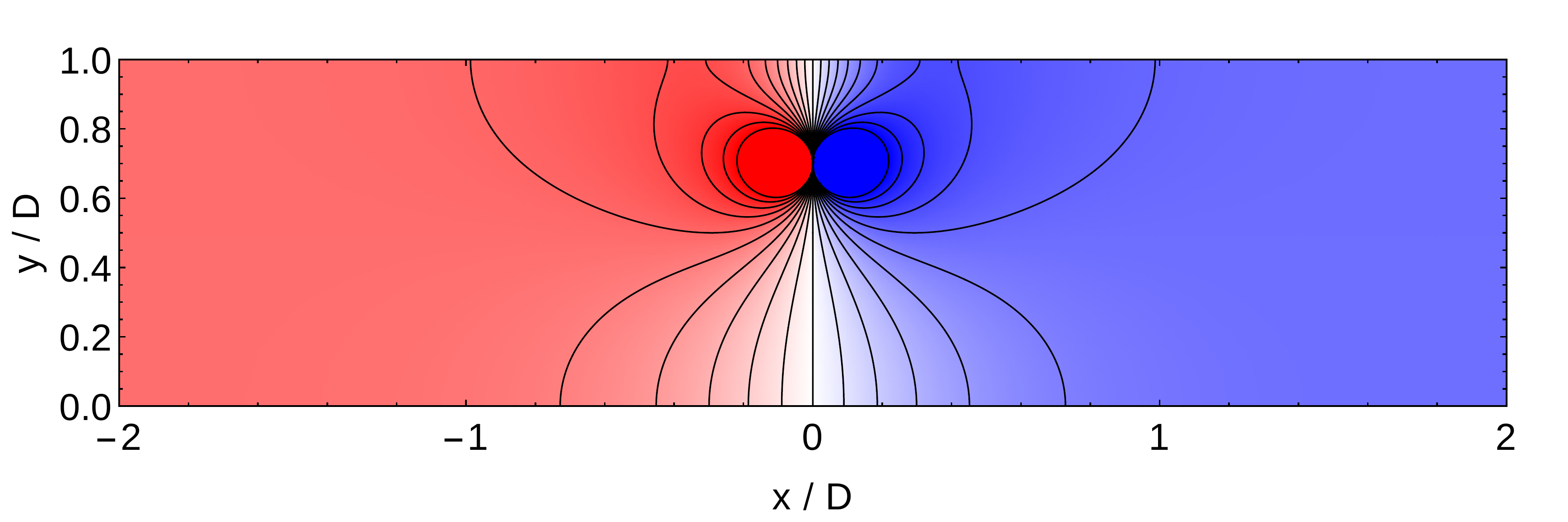}};
\node[scale=0.35] at (3.95,\yoffset+\yoffsetl) {\includegraphics[width=0.03\textwidth]{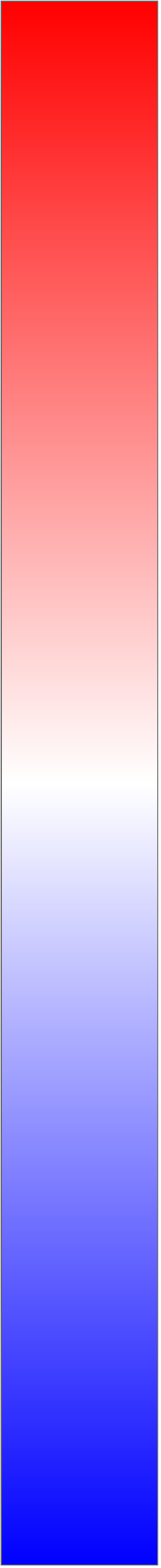}};
\node[scale=0.8] at (4.4,\yoffset-0.9+\yoffsetl) {$-\pi$};
\node[scale=0.8] at (4.3,\yoffset+\yoffsetl) {$0$};
\node[scale=0.8] at (4.4,\yoffset+0.9+\yoffsetl) {$\pi$};
\node at (-3,\yoffset+0.6+\yoffsetl) {(a)};
\node[scale=0.95] at (0,2*\yoffset) {\includegraphics[width=0.47\textwidth]{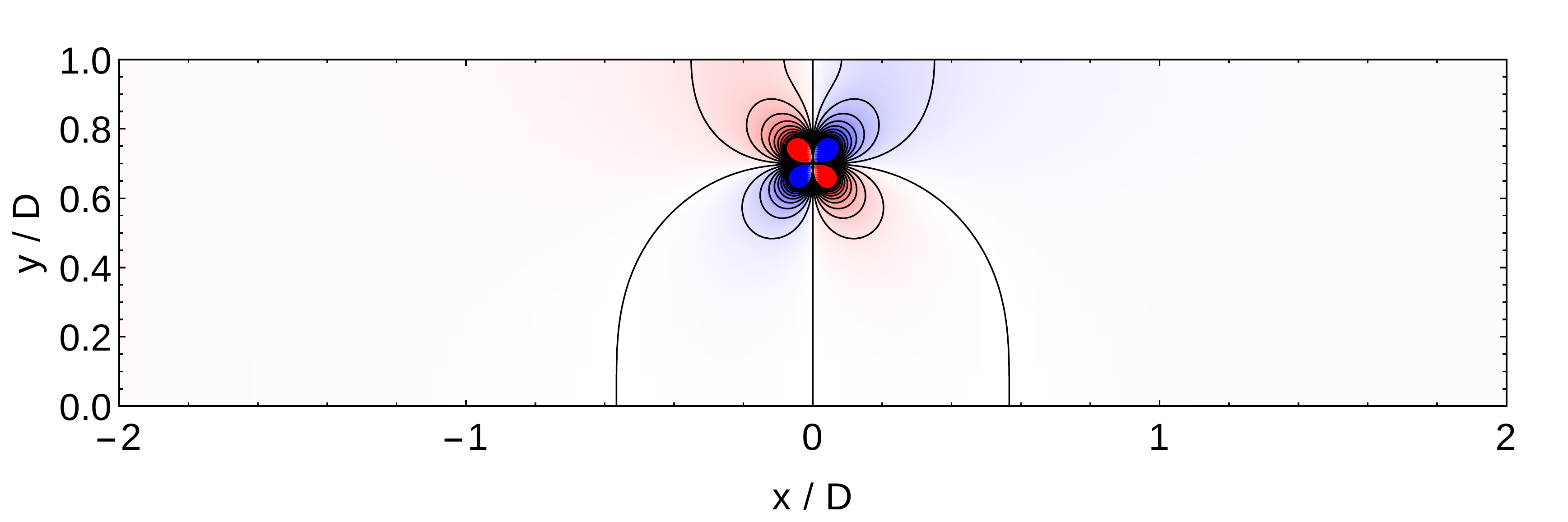}};
\node[scale=0.35] at (3.95,2*\yoffset+\yoffsetl) {\includegraphics[width=0.03\textwidth]{1l_SI.pdf}};
\node[scale=0.8] at (4.4,2*\yoffset-1.0+\yoffsetl) {$-\pi/10$};
\node[scale=0.8] at (4.3,2*\yoffset+\yoffsetl) {$0$};
\node[scale=0.8] at (4.4,2*\yoffset+0.9+\yoffsetl) {$\pi/10$};
\node at (-3,2*\yoffset+0.6+\yoffsetl) {(b)};
\node[scale=0.95] at (0,3*\yoffset) {\includegraphics[width=0.47\textwidth]{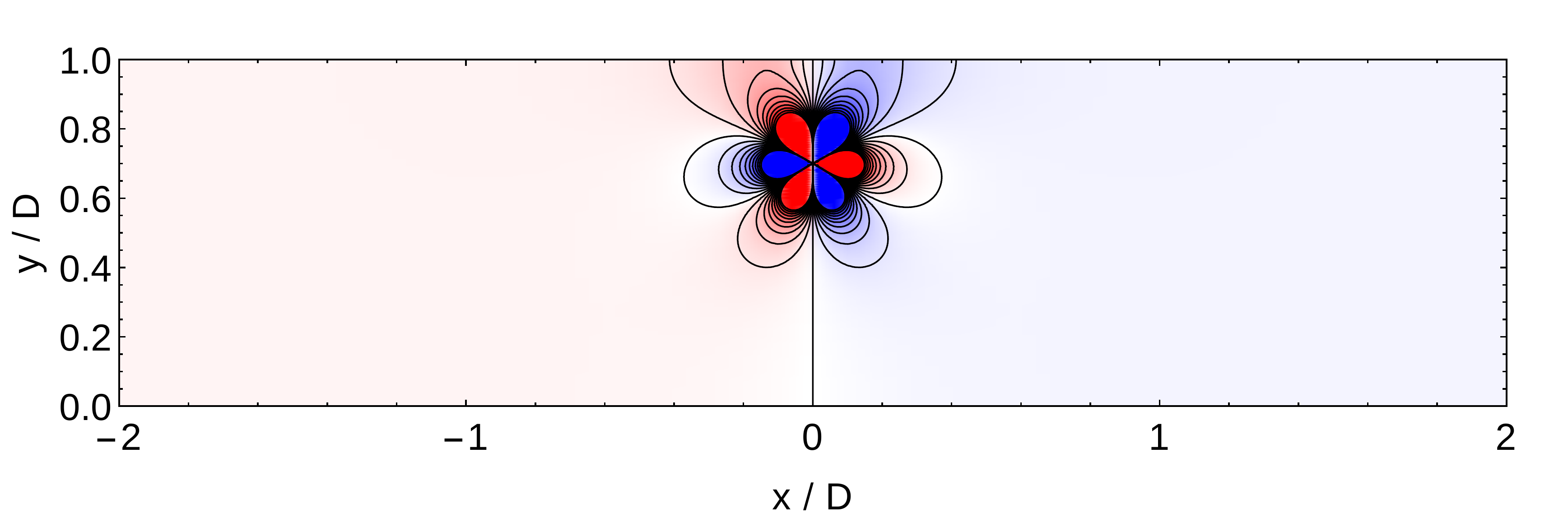}};
\node[scale=0.35] at (3.95,3*\yoffset+\yoffsetl) {\includegraphics[width=0.03\textwidth]{1l_SI.pdf}};
\node[scale=0.8] at (4.5,3*\yoffset-1.0+\yoffsetl) {$-\pi/1000$};
\node[scale=0.8] at (4.3,3*\yoffset+\yoffsetl) {$0$};
\node[scale=0.8] at (4.5,3*\yoffset+0.9+\yoffsetl) {$\pi/1000$};
\node at (-3,3*\yoffset+0.6+\yoffsetl) {(c)};

\node[scale=0.95] at (9,\yoffset) {\includegraphics[width=0.47\textwidth]{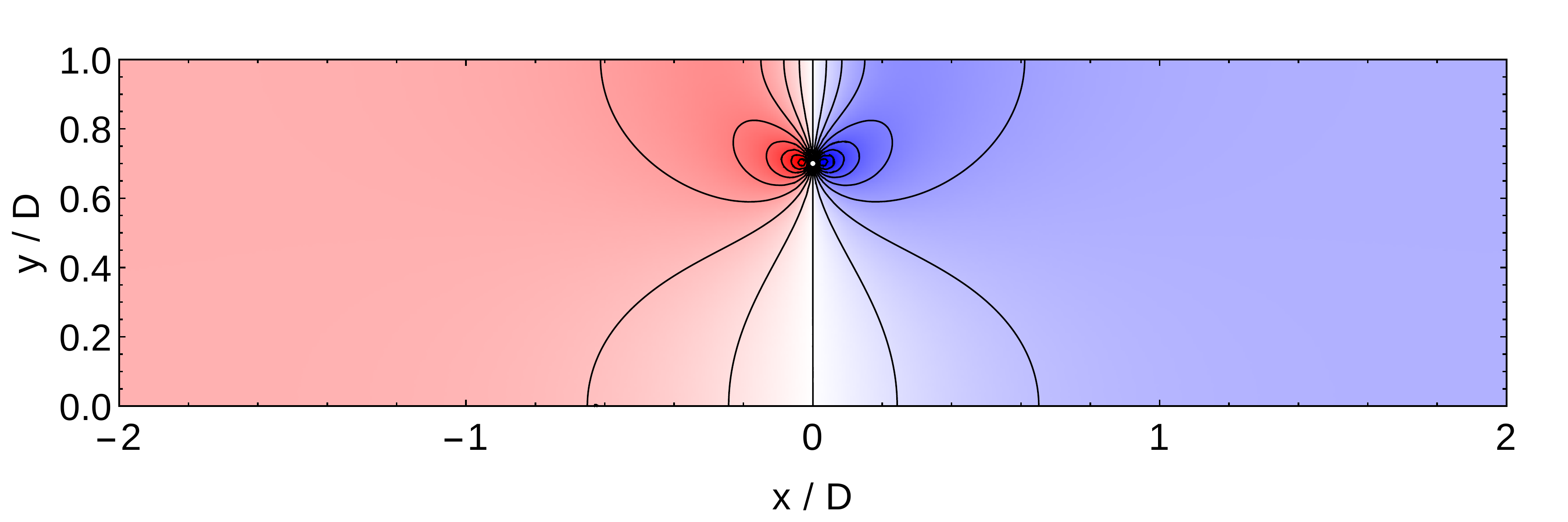}};
\node[scale=0.35] at (9+3.95,\yoffset+\yoffsetl) {\includegraphics[width=0.03\textwidth]{1l_SI.pdf}};
\node[scale=0.8] at (9+4.4,\yoffset-0.9+\yoffsetl) {$-80$};
\node[scale=0.8] at (9+4.3,\yoffset+\yoffsetl) {$0$};
\node[scale=0.8] at (9+4.4,\yoffset+0.9+\yoffsetl) {$80$};
\node at (9-3,\yoffset+0.6+\yoffsetl) {(d)};

\node[scale=0.95] at (9,3*\yoffset) {\includegraphics[width=0.47\textwidth]{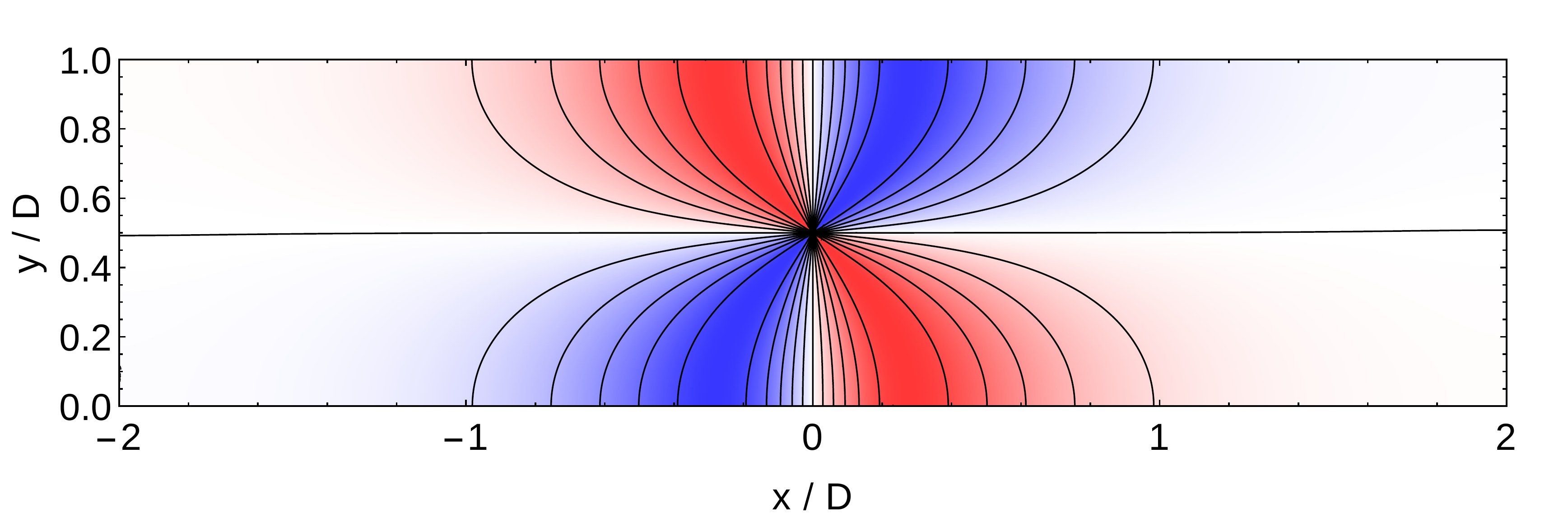}};
\node[scale=0.35] at (9+3.95,3*\yoffset+\yoffsetl) {\includegraphics[width=0.03\textwidth]{1l_SI.pdf}};
\node[scale=0.8] at (9+4.4,3*\yoffset-1.0+\yoffsetl) {$-\pi$};
\node[scale=0.8] at (9+4.3,3*\yoffset+\yoffsetl) {$0$};
\node[scale=0.8] at (9+4.4,3*\yoffset+0.9+\yoffsetl) {$\pi$};
\node at (9-3,3*\yoffset+0.6+\yoffsetl) {(f)};

\node[scale=0.95] at (9,2*\yoffset) {\includegraphics[width=0.47\textwidth]{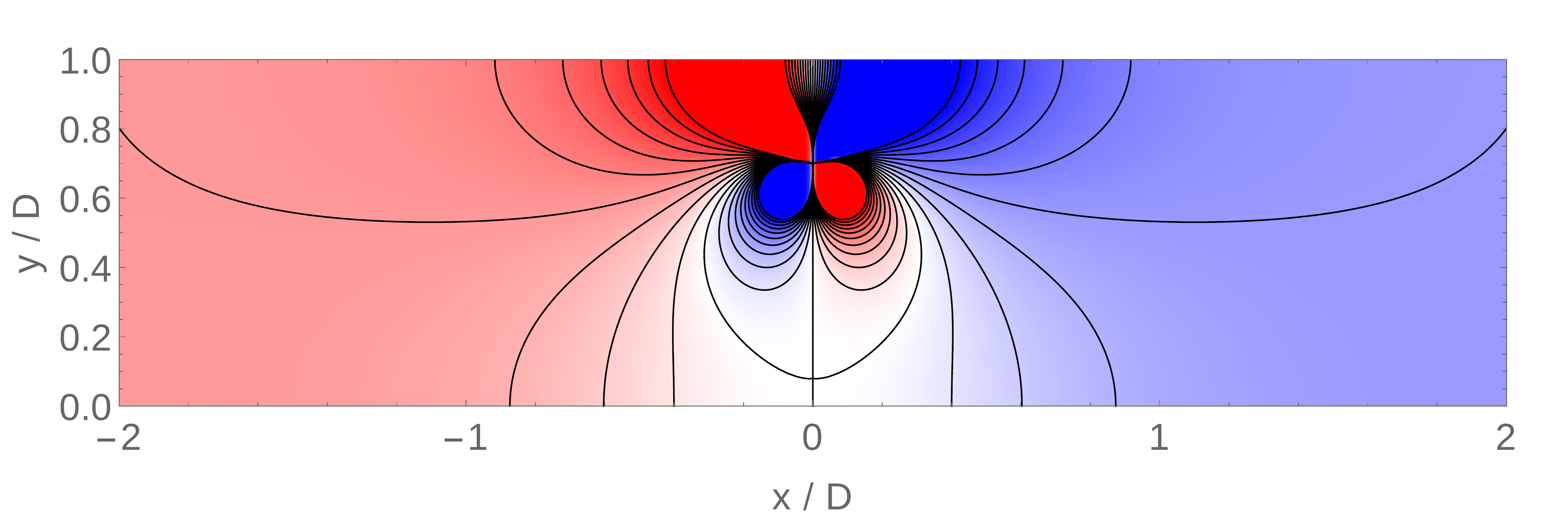}};
\node[scale=0.35] at (9+3.95,2*\yoffset+\yoffsetl) {\includegraphics[width=0.03\textwidth]{1l_SI.pdf}};
\node[scale=0.8] at (9+4.4,2*\yoffset-1.0+\yoffsetl) {$-4\pi$};
\node[scale=0.8] at (9+4.3,2*\yoffset+\yoffsetl) {$0$};
\node[scale=0.8] at (9+4.4,2*\yoffset+0.9+\yoffsetl) {$4\pi$};
\node at (9-3,2*\yoffset+0.6+\yoffsetl) {(e)};

\end{tikzpicture}
\caption{\label{fig:multipole}
(a) $S_1^\mathrm{approx}/(q^3 \xi^2/D^2)$ to leading order i.e. no $R$-dependence, (b) next-to-leading order $R^2/D^2$, (c) and order $R^4/D^4$. (d) Numerical evaluation of $S_1/(q^3 \xi^2/D^2)$ with an excision ($R/D = 0.01$) at the vortex core. Note the (divergent) scale from having a small $R$. (e) Renormalised, cut-off independent $S_1/(q^3 \xi^2/D^2)$. In (a)-(e), $h/D = 0.7$. (f) Renormalised, cut-off independent $S_1/(q^3 \xi^2/D^2)$ at $h = D/2$.}
\end{figure*}

The Poisson equation~\eqref{eqn:s111} can be solved with the help of a Green's function that obeys the Neumann boundary conditions of the channel and satisfies  $\nabla^2 G(\textbf{r},\textbf{r}^\prime) = +\delta^{(2)}(\textbf{r}-\textbf{r}^\prime)$. The Neumann Green's function for the channel satisfying $\partial_n G = 0$ on the channel walls along $y = 0$ and $y = D$, and $\partial_n G = 1/(2D)$ on the walls at $|x| \to \infty$, where $\partial_n$ is the unit outward normal derivative, is~\cite{barton1989elements}
\begin{equation}
\begin{split}
G(\textbf{r},\textbf{r}^\prime) &= \frac{1}{4\pi} \ln \left\lbrace \left[\sinh ^2\left(\frac{\pi  (x-x^\prime)}{2 D}\right)+\sin ^2\left(\frac{\pi  (y-y^\prime)}{2 D}\right)\right] \right. \\ 
&\times \left. \left[\sinh ^2\left(\frac{\pi  (x-x^\prime)}{2 D}\right)+\sin ^2\left(\frac{\pi  (y+y^\prime)}{2 D}\right)\right]\right\rbrace.
\end{split}
\end{equation}
The solution is then found by integrating the Green's function with the source term 
\begin{equation} \label{eq:S1int}
S_1(\textbf{r}) = \int_{\Omega \setminus B_{\boldsymbol{\rho}}^R} -\frac{1}{n_0} \nabla \cdot (n_1(\textbf{r}^\prime) \nabla S_0(\textbf{r}^\prime)) G(\textbf{r},\textbf{r}^\prime) \mathrm{d}^2\textbf{r}^\prime ,
\end{equation}
where we have regularised the otherwise divergent integral by excising a disk $B_{\boldsymbol{\rho}}^R$ with radius $R$.  This integral can be solved numerically, and a renormalisation procedure that allows $R$ to be taken to zero is outlined below. Further analytical progress is made noting that 
$\nabla^2 S_0 =0$ in the integration domain and writing $G \nabla n_1 \cdot \nabla S_0 = G \nabla \cdot (n_1 \nabla S_0) =  \nabla \cdot \left[G n_1 \nabla S_0 \right] - n_1 \nabla S_0 \cdot \nabla G$:
\begin{equation}
\label{eqn:S1_int}
S_1(\textbf{r}) =  \int_{\Omega \setminus B_{\boldsymbol{\rho}}^R} -\frac{1}{n_0}\left\lbrace  \nabla \cdot \left[G n_1 \nabla S_0 \right] - n_1 \nabla S_0 \cdot \nabla G\right\rbrace \mathrm{d}^2\textbf{r}^\prime. 
\end{equation}
The first term on the right hand side can be analytically evaluated as a boundary term on $\partial B_{\boldsymbol{\rho}}^R$, which is the circle where $|\textbf{r}^\prime -\boldsymbol{\rho} | = R$. Setting $\boldsymbol{\rho} = (0,h)$, the position of the vortex, we have $(x^\prime)^2 + (y^\prime - h)^2 = R^2$, so that $x^\prime = R \cos(\theta)$ and $y^\prime = h+R\sin(\theta)$, where $\theta$ is measured at the vortex. We have to integrate over all $\theta$, and obtain as an approximation for $S_1$ that ignores the second term on the right hand side of Eq.~\eqref{eqn:S1_int}
\begin{widetext}
\begin{equation}
\label{eqn:S1}
\begin{split}
S_1^\mathrm{approx}(\textbf{r}) &=  -\frac{\pi ^2 q^3    }{4 }\frac{\xi^2}{D^2}\frac{ \cot \left(\frac{\pi  h}{D}\right) \sinh \left(\frac{\pi  x}{D}\right) \left[\cosh \left(\frac{\pi  x}{D}\right)-\cos \left(\frac{\pi  h}{D}\right) \cos \left(\frac{\pi  y}{D}\right)\right]}{\left[\cos \left(\frac{\pi  (h-y)}{D}\right)-\cosh \left(\frac{\pi  x}{D}\right)\right] \left[\cosh \left(\frac{\pi  x}{D}\right)-\cos \left(\frac{\pi  (h+y)}{D}\right)\right]} + \mathcal{O}\left( \frac{R^2}{D^2} \right) ,
\end{split}
\end{equation}
\end{widetext}
where $\textbf{r} = (x,y)$.
The $(R/D)^2$ series corresponds to a `multipole-like' expansion (Fig.~\ref{fig:multipole}). We take the limit $R \to 0$ so that in effect we are excising a point.

\subsection{Renormalisation}
At first sight, the integral \eqref{eq:S1int} for $S_1$ is divergent. However, as we outline here, our model is renormalisable, and the physical properties of the solitonic vortex can be meaningfully assigned.

First, we regularise the divergence by excising the disk $B_{\boldsymbol{\rho}}^R$. The analytically solvable term $S_1^\mathrm{approx}$ equates to a multipole-like expansion in the cut-off $R$ (Fig.~\ref{fig:multipole}a,b,c), while the term that was ignored in Eq.~\eqref{eqn:S1_int} is divergent as $R \to 0$. More specifically, it integrates to a divergent dipole term (Fig.~\ref{fig:multipole}d), and an $R$-independent quadrupole term (Fig.~\ref{fig:multipole}f). The dipole terms from both the analytically solvable term and the numerically evaluated term vanish at $h=D/2$, and $S_1$ takes the form of a quadrupole field there (Fig.~\ref{fig:multipole}f). Since the quadrupole field has no $R$-dependence in the full $S_1$ (it has $R^2$ dependence only in the analytically solvable term $S_1^\mathrm{approx}$), the integral for $S_1$ is already renormalised and indeed gives an identical result as the full renormalisation procedure.

To renormalise the divergences, we expand the integrand in Eq.~\eqref{eqn:S1_int} around the vortex position, and add counter-terms to cancel the terms in the expansion with negative powers of $\tilde{r}$, where $\tilde{r}$ is the distance from the vortex core. We have
\begin{equation}
-\frac{1}{n_0} \nabla \cdot (n_1(\textbf{r}^\prime) \nabla S_0(\textbf{r}^\prime)) G(\textbf{r},\textbf{r}^\prime) = \frac{a_3}{\tilde{r}^3} + \frac{a_2}{\tilde{r}^2} + \frac{a_1}{\tilde{r}} + \ldots,
\end{equation}
and so we need three counter terms. Here $a_{1-3}$ are analytic expressions that depend on $\textbf{r}$ and $\textbf{r}^\prime$. Then, we take $R\to 0$ and obtain finite values for the remaining integrals. By doing this we find that $S_1$ is in general a combination of a dipole term and a quadrupole term (Fig.~\ref{fig:multipole}e). As discussed, only the finite quadrupole term survives at $h = D/2$. Although the integral  \eqref{eq:S1int} for $S_1$ formally diverges when $h\neq D/2$, the underlying physical model is thus renormalisable with only a finite number of counter terms required.

\bibliography{Fermi_Gas,Solitons1,references}

\end{document}